\documentclass[twocolumn]{aastex63}
\hypersetup{linkcolor=red,citecolor=green,filecolor=cyan,urlcolor=magenta}
\begin{document}
\newcommand{\vdag}{(v)^\dagger}
\newcommand\aastex{AAS\TeX}
\newcommand\latex{La\TeX}
\newcommand{\ct}{$^{13}$C}[
\newcommand{\ctb}{$^{13}$C~}
\newcommand{\cd}{$^{12}$C}
\newcommand{\cdb}{$^{12}$C~}
\newcommand{\cta}{$^{13}$C($\alpha$,n)$^{16}$O}
\newcommand{\ctab}{$^{13}$C($\alpha$,n)$^{16}$O~}
\newcommand{\nean}{$^{22}$Ne($\alpha$,n)$^{25}$Mg}
\newcommand{\neanb}{$^{22}$Ne($\alpha$,n)$^{25}$Mg~}
\newcommand{\fe}{$^{56}$Fe}
\newcommand{\feb}{$^{56}$Fe~}
\newcommand{\fes}{$^{60}$Fe}
\newcommand{\fesb}{$^{60}$Fe~}
\newcommand{\al}{$^{26}$Al}
\newcommand{\alb}{$^{26}$Al~}
\newcommand{\oq}{\textquotedblleft}
\newcommand{\cq}{\textquotedblright}
\newcommand{\cqb}{\textquotedblright~}
\newcommand{\ms}{M$_{\odot}$}
\newcommand{\msb}{M$_{\odot}$~}
\received{}
\revised{}
\accepted{}
\submitjournal{ApJ}
\shorttitle{MHD-induced $s$-processing}
\shortauthors{Busso et al.}
\title{$s$-Processing in AGB Stars Revisited. III. \\
Neutron captures from MHD mixing at different metallicities and observational constraints}
\correspondingauthor{Maurizio Busso}
\email{maurizio.busso@unipg.it, maurizio.busso@pg.infn.it}

\author{Maurizio Busso}
\affiliation{Department of Physics and Geology, University of Perugia,  Via A. Pascoli snc, I-06123 Perugia, Italy} 
\affiliation{INFN, section of Perugia, Via A. Pascoli snc, I-06123 Perugia, Italy}

\author{Diego Vescovi}
\affiliation{Gran Sasso Science Institute, Viale Francesco Crispi, 7, I-67100 L’Aquila, Italy}
\affiliation{INFN, section of Perugia, Via A. Pascoli snc, 06123 Perugia, Italy}
\affiliation{INAF, Observatory of Abruzzo, Via Mentore Maggini snc, I-64100 Collurania, Teramo, Italy}

\author{Sara Palmerini}
\affiliation{Department of Physics and Geology, University of Perugia, Via A. Pascoli snc, I-06123 Perugia, Italy} 
\affiliation{INFN, section of Perugia, Via A. Pascoli snc, I-06123 Perugia, Italy}

\author{Sergio Cristallo}
\affiliation{INAF, Observatory of Abruzzo, Via Mentore Maggini snc, I-64100 Collurania, Teramo, Italy}
\affiliation{INFN, section of Perugia, Via A. Pascoli snc, I-06123 Perugia, Italy}

\author{Vincenzo Antonuccio-Delogu}
\affiliation{INAF, Catania Astrophysical Observatory, Via S.Sofia 79, I-95129 Catania, Italy}

\begin{abstract}
We present post-process neutron-capture computations for Asymptotic Giant Branch (AGB) stars of 1.5$-$3 \msb and metallicities $-1.3\leq$[Fe/H]$\leq$0.1. The reference stellar models are computed with the FRANEC code, using the Schwarzschild’s criterion for convection; motivations for this choice are outlined. We assume that MHD processes induce the penetration of protons below the convective boundary, when the Third
Dredge Up occurs. There, the \ctb $n$-source can subsequently operate, merging its effects with
those of the \neanb reaction, activated at the temperature peaks characterizing AGB stages. This 
work has three main scopes. i) We provide a grid of abundance yields, as produced through our MHD 
mixing scheme, uniformly sampled in mass and metallicity. From it, we deduce that 
the solar $s$-process distribution, as well as the abundances in recent stellar populations,
can be accounted for, without the need of the extra primary-like contributions suggested in 
the past.   ii) We formulate analytic expressions for the 
mass of the \ct-pockets generated, in order to allow easy verification of 
our findings. {iii) We compare our results with observations of
evolved stars and with isotopic ratios in presolar SiC grains, also noticing how some 
flux tubes should survive turbulent disruption, carrying C-rich 
materials into the winds even when the envelope is O-rich. This wind phase is
approximated through the {\it G-component} of AGB $s$-processing}. We conclude that 
MHD-induced mixing is adequate to drive slow $n$-capture phenomena accounting 
for observations; our prescriptions should permit its inclusion into current stellar 
evolutionary codes.
\end{abstract}

\keywords{Nucleosynthesis, $s$-process --- Stars, evolution --- Stars, abundances --- Galaxies, chemical evolution}
\vskip 0.3cm
\section{Introduction: setting the stage} \label{sec:intro}
Stars are made of plasmas, in which physical conditions range over several orders 
of magnitude in pressure, temperature and density. In them, many hydro-dynamical and 
magneto-hydro-dynamical processes occur in variable and complex ways, 
characterized by micro- and macro-turbulence phenomena with Reynolds' numbers well 
beyond the limits experimentally studied in terrestrial laboratories \citep{tsuji} and 
actually also beyond our capability of detailed, quantitative modeling.
Evolutionary computations can only ascertain that traditional one-dimension models 
are largely insufficient to account for short- and long-term
processes of stirring and mixing \citep{stan6} and often limit themselves to simulate
schematically the layers affected by pure convection, using the mixing length 
theory or some other simplified approaches \citep[see][and references 
therein]{sal1, sal2}. They can also address in similar ways 
the regions of semi convection, distinguishing between the two mixing 
schemes through the Schwarzschild's and Ledoux's criteria for stability, 
both based on simple polythropic approaches \citep{chan}. In modern 
computations, they may include also some (but not all) of the effects 
induced by rotation \citep[see e.g.][and references therein]{rot3, rot1, rot2,
rot5, stan7}. These limits, in existing efforts, are unavoidable and make clear that the 
real behavior of stellar plasmas is more complex than 
our basic descriptions. Even for the Sun, a large family of different 
dynamical processes induce variations in the structure, hence in the irradiance, 
over time scales ranging from minutes to billions of years \citep{kopp}.
It is therefore expected that also in the advanced stages of their life, during and 
after the ascent to the Red Giant Branch (RGB), stars sharing the same evolutionary scheme 
of the Sun experience mass and momentum transfer at different speed \citep{jerry1, char}. 
These are the stellar objects of low and intermediate mass (LMS and IMS: i.e. those in 
the mass ranges 1 to 3 \msb and 3 to about 8 \ms, respectively). The exercise of imagining, for them, 
which circulation or diffusion mechanisms might be at play is required at least by 
the need of reproducing isotopic and elemental observations that cannot be explained 
by usual models with pure convection \citep{b+9, n+03, b+10, kl14}.  Recent research 
has focused on many such mechanisms, from relatively fast dynamical \citep{dt, pig, bat1, bat2} 
and magneto-hydro-dynamical \citep{b+07, nor} processes with speed up to several m/sec, 
all the way down to various  forms of slow (less than 1 cm/sec) diffusive mixing 
\citep{cz8,egg1,egg2,stan1}.

\begin{figure}[t!!]
\includegraphics[width=0.85\columnwidth]{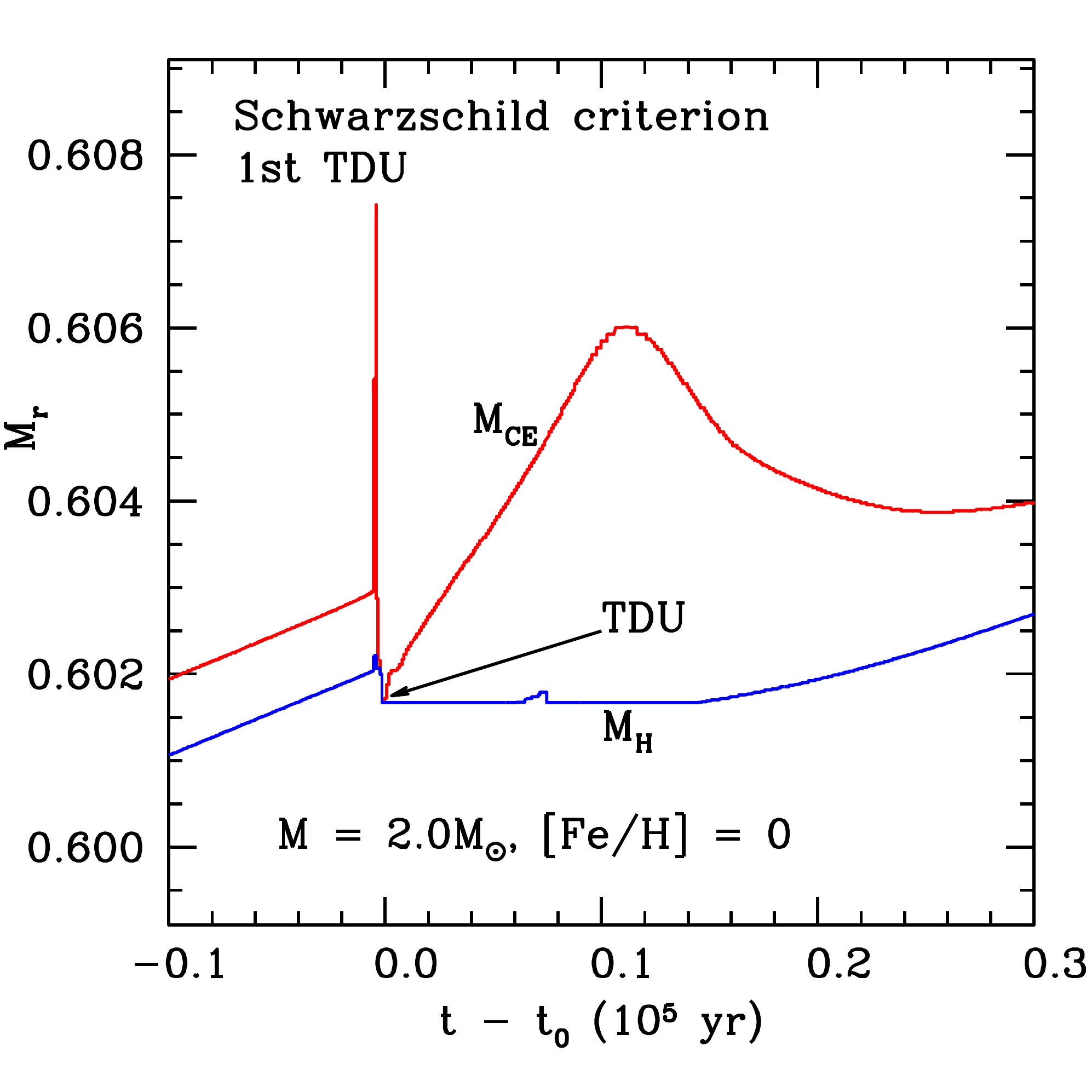}
\caption{The first occurrence of the envelope penetration in a Third Dredge Up episode, 
computed with the FRANEC code for a 2 \msb star of solar composition. The figure 
shows in red the innermost border of the convective envelope ($M_{\rm CE}$) and in blue 
the position of the H/He interface ($M_{\rm H}$), with its characteristic minimum 
({\it post-flash dip}) before H-burning restarts. The parameter $t_0$ is the stellar 
age at the moment of the TDU episode shown. Note how, of the rather long duration of 
the post-flash dip ($\sim 10^4$ yr), only a very short fraction is occupied by 
TDU ($\sim$ 100 yr).  \label{fig:tdu}}
\end{figure}

It is now ascertained that some abundance observations provide constraints 
at least on the velocity, and possibly on the nature, of the dynamical mixing phenomena 
occurring \citep{her2, b+07, pal2, den1, liu}. This is so in particular
for Li \citep{cl10, pal2} and for the enrichment in neutron-rich elements
\citep{cris0, cris1, t+16, cris2} occurring in the final evolutionary stages, 
approaching the RGB  asymptotically (AGB phases). Indeed, it has been known for 
more than 30 years now that the bulk of neutron fluxes for the required 
nucleosynthesis episodes is produced in \ctb reservoirs \citep{gal0, 
gal1, ar9, ka11}, formed at the recurrent penetration of envelope convection 
into the He-rich buffer of the star, called the Third Dredge Up (TDU). This
last phenomenon has now been observed to occur over the whole range of the masses
here considered and down to about 1 \msb stars \citep{shetye1}.
The mentioned suggestions  on $s$-processing  have been verified in some detail 
directly on the observations of carbon stars \citep{abia1, abia2, abia3}. Formation 
of the \ctb reservoir follows 
the retreat of shell convective instabilities developing in the He-layers 
\citep{stra1, her1} and requires that an important fraction of the He-rich buffer 
(hereafter {\it He-intershell} zone) 
be swept by the penetration of protons from the envelope \citep[][hereafter paper I]{t+14}, 
within the time interval over which the bottom border of envelope convection reaches 
its maximum downward expansion (see Figure \ref{fig:tdu}, for a star of solar metallicity
\footnote{We {remind the reader} that the heavy element
content of a star relative to the Sun, i.e. its {\it metallicity}, is commonly indicated in logarithmic notations,
with the parameter ${\rm [Fe/H]} = Log($X$(Fe)/$X$(H))_{star} - Log($X$(Fe)/$X$(H))_{\odot}$}).

The limited duration in time of
the TDU phenomenon ($\simeq$ 100 yr) plays in favor of rather fast mixing mechanisms
(of the order of a few m/sec). This can be e.g. achieved in the case of the unimpeded 
buoyancy of magnetic flux tubes, which may occur given the specific polythropic 
structure of high index ($n \ge 3$) and the fast density decline 
($\rho \propto r^k$, with $k << -1$), prevailing in the radiative
layers below the convective envelope, which provide an exact analytical solution
to MHD equations in the form of free, accelerating expansion \citep{nb}. 
We notice how the alternative to look for exact solutions in a simplified
geometry would be that of performing detailed 3D numerical simulations, an approach
that has indeed seen important attempts \citep[see e.g.][and references therein]{stan2}, but is much more difficult to
implement in general.

The first neutron-capture nucleosynthesis computations presented in the MHD scenario 
\citep[][]{t+16} demonstrated that the peculiar profile of 
\ctb in the reservoirs thus formed was suitable for reproducing detailed isotopic
patterns in presolar grains that could not be fitted otherwise, in agreement 
with previous indications by \citet{liu}. This was shown to be possible in
a framework that could also mimic the solar $s$-process distribution; it was also suggested 
how extrapolations of that model could satisfy other observational requirements.
In this contribution we want to extend the work presented in 
\citet[][hereafter paper II]{t+16} by computing the formation of MHD-induced 
\ctb pockets over a wide range of metallicities and for FRANEC evolutionary models 
of 1.5 to 3 \ms. In so doing, we also provide analytic fits to the extension 
of the \ct-pockets, in order to make our results easily reproduced by others. 
We outline our approach and assumptions, as well as the mentioned analytic fits,
together with general results as a function of mass and metallicity,
in Section 2. Subsequently, in Section 3 we discuss how
our results can be used,  once weighted in mass and time using a common choice 
for the Initial Mass Function (IMF) and a Star Formation Rate (SFR) taken 
from the literature, in accounting for the abundances gradually built by Galactic
evolution and now observed in stars of the Galactic disk, from our Solar System to 
the most recent stellar populations of young open clusters. This is a synthetic 
anticipation from a more extended work of chemo-dynamical Galactic evolution we are pursuing.
Finally, in Section 4 we compare the abundance distributions produced 
by our models at the surface of evolved stars with some of the observational 
constraints available either from actual AGB and post-AGB stars of various 
families, or from the isotopic analysis of trace elements in presolar grains.
The series of comparisons presented in this paper have the final goal to
guarantee that our predictions can be safely and robustly verified. As a 
consequence of these checks, our mixing prescriptions have now 
been delivered for direct inclusion into full stellar models of the FUNS 
series \citep{cris3}: preliminary results of such an inclusion were presented 
in a recently published paper \citep{diego}. 

\section{Modeling the TP-AGB evolution and its nucleosynthesis}

\subsection{The stellar models and the proton mixing}
In paper II we implemented the exact solution 
of MHD equations, as found by \citet{nb} in the form of free buoyancy 
for magnetic flux tubes, and on that basis we computed neutron-capture 
nucleosynthesis in a star of 1.5 \msb of a metallicity slightly lower 
than solar. Then we adopted the ensuing model as a 
proxy for stars in the mass range from 1 to 3 \ms, at Galactic-disk 
metallicities. This preliminary, simplified extrapolation was 
motivated by previous suggestions advanced in \citet{mai1, mai2} and in 
paper I. In these works, starting from 
parameterized extensions of the \ctb pocket, it had actually been 
found that the nuclear yields of a star characterized by parameters 
(mass, initial metallicity, extension of the \ctb reservoir) very 
similar to those of the model shown in paper II would provide a 
reasonable approximation to the average ones in the Galactic disk,
thus mimicking the enrichment of the Solar System and of 
recent stellar populations in neutron-rich nuclei.

We want now to complete the job by estimating, for a wide
metallcity range ($-1.3 \leq$ [Fe/H] $\leq 0.1$) and for three reference 
stellar masses (1.5, 2.0 and 3.0 \ms, computed ad-hoc with the FRANEC 
evolutionary code) the extensions of the \ctb pockets formed in the
hypotheses of paper II. In particular, we assumed that proton penetration from the envelope 
into the He-rich layers, at every TDU episode, occurs as a consequence of the 
activation of a stellar dynamo, with the ensuing buoyancy into the envelope
of highly magnetized structures. These last push down poorly magnetized material,
forcing it into the radiative He-rich layers. 

As mentioned previously, the stellar models were computed with the FRANEC evolutionary
code, which uses the Schwarzschild's criterion for convection: for a description of 
the physical assumptions characterizing the code, see e.g. \citet{str}. 

It is today generally recognized that some form of extension of the convective border with 
respect to a pure Schwarzschild's limit is needed \citep{frey}; see on this point the 
discussion by \citet{vent} and the references cited therein. However, as we want 
here to look for how the \ctb $n$-source is formed in MHD-driven mechanisms,
our approach must be that of attributing any extension of such a border to
magnetic effects, without the admixture of different schemes, each based on
its own free parameters, which would make the disentangling of different effects
ambiguous. Only after this work is completed we are authorized to check for 
possible changes induced by a different treatment of the convective extension, as 
is in fact done in \citet{diego}. 
Originally,
our models adopted a Reimers' criterion for mass loss, with the parameter $\eta$
set to 1.0 for 1.5 and 2.0 \msb models, and to 3 for 3.0 \msb models. In making
post-process computations for nucleosynthesis, we adopted instead the
more efficient mass loss rates of the FRUITY repository. The rate of mass loss through stellar 
winds remains in any case largely unknown; this fact introduces imporant uncertainties
on the composition of the stellar envelopes \citep{stan5}. 

In this context, it will be necessary to compute not only the average abundances gradually formed in the 
envelopes, but also those of the He-shell material cumulatively transported by TDU episodes. This
represents an $s$-process-enhanced, C-rich phase averaged over the efficiency of 
mixing, which is called, since many years \citep{zinner}, the {\it G component}. {In our 
models, the $G$-component carries abundances very similar to those of flux tubes 
that, due to strong magnetic tension, were to survive destruction by turbulence in the 
convective envelope, opening later in the wind, as occurring in the Sun \citep{pinto}.
There is actually some support in the current literature to the existence of such 
magnetized wind structures in evolved  stars \citep{sk03, sab15, ros91, ros92}. 
Such blobs would maintain an unmixed C- and $s$-process
rich composition, typical of the He-intershell zone, even when the rest of the envelope 
is O-rich.  In our models, using the $G$-component to approximate the abundances of 
this wind phase is feasible because of the high neutron fluences generated in each 
pulse-interpulse cycle, which usually allow the effects of the most recent nucleosynthesis 
episode to dominate over the previous ones.} The relevance of this approximation will 
become clear in considering the isotopic admixtures of presolar SiC grains enriched in 
$s$-process elements (see section 4.1).

\subsection{The \ctb pocket and the ensuing nucleosynthesis}

In the above approach, the extension and profile of any proton reservoir formed 
are computed according to the formulation of paper II (see there equations from 14 to 17), 
after verifying that the required conditions, stated in \citet{nb}, are satisfied.
The occurrence of the proper physical conditions was ascertained as follows.

\begin{figure*}[t!!]
\center{
\includegraphics[width=0.6\linewidth]{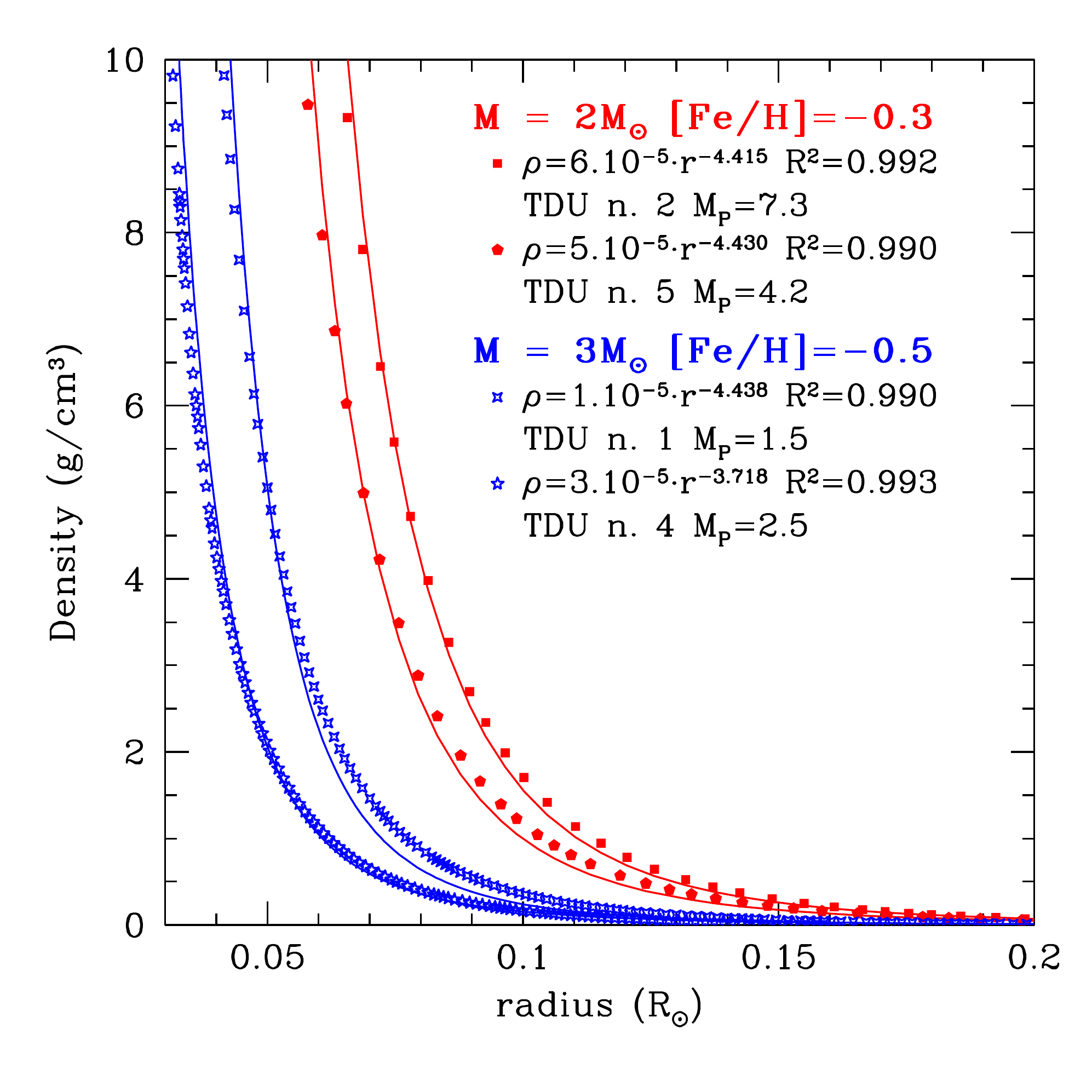}
\caption{Examples of how the density distributions below the formal convective 
envelope bottom (defined by the Schwarzschild's criterion) agree with the
requirement of being exact power laws of the radius with large ($k < -3$)
negative exponents. The cases shown are from the second and fifth TDU
occurrence in a 2 \msb star of half-solar metallicity and from the first and fourth
TDU occurrence in a 3 \msb star of one-third-solar metallicity. The fitting
power laws are shown, with their regression coefficients. Also indicated are the
resulting masses $M_p$ of the proton reservoirs, expressed in units of
10$^{-3}$ \ms. In models for the largest stellar masses considered (3 \ms) the 
validity of the solution is generally limited to layers considerably thinner 
than for lower mass models.}\label{fig:pockets}}
\end{figure*}

i) The MHD solution found in \citet{nb} and expressed in equation (5)
of paper II ($\rho(r) \propto r^{k}$, with $k < -1$) was verified 
on the model structures computed with the stellar code, for every TDU occurrence
(in the He-rich layers $k$ turns out to be always lower than $-3$). 
Since the condition on $k$ {derives} from an exact solution, 
we expect that the regression coefficients be close to 1. They are always
larger than 0.98 over about half of the intershell region (in mass). In the
layers of that zone where protons penetrate according to the equations of
paper 2 they are even larger, reaching typical values as indicated here
in Figure 1.

ii) Over the layers thus selected, we computed the kinematic
viscosity $\eta$ from the approach by \citet{scheck}, as:
$$
\eta(r) \propto \frac{v_{th}(r)}{n(r)\sigma(r)} \eqno(1)
$$
where $v_{th}(r)$ is the local thermal velocity and $\sigma(r)$ is the ionic 
cross section, $\pi l(r)^2$, $l$ being proportional to the DeBroglie's wavelength,
$l \propto h/(m {v_{th})}$. From this, we estimated the dynamical viscosity:
$$
\mu(r) = \eta(r) \rho(r) \eqno(2)
$$
We then assumed from \citet{spiz} and  \citet{scheck} the value of the magnetic
Prandtl number, as:
$$
P_m(r) = \frac{\eta}{\nu} \simeq 10^{-5} \frac{T(r)^4}{n(r)} \eqno(3)
$$
(where $\nu$ is the magnetic viscosity) and verified that values of $P_m$ were
always much larger than unity (they turned out to
be always larger than 10 over the selected layers). iii) We also requested
that the third condition posed by \citet{nb} were verified, namely that
the region of interest had a {\it low} value for the dynamical viscosity
(as defined in equation 2), i.e. values of $\mu$ much smaller than in more 
internal regions. This condition was again
found to hold easily, due to the steep growth of the density in the innermost
layers of the He-intershell zone. A few examples of the fits obtained under
the above conditions, with the resulting masses for the proton reservoirs are
shown in Figure \ref{fig:pockets}. 

\begin{deluxetable*}{c|ccc|ccc|ccc}
\tablenum{1}
\tablecaption{The coefficients of the parabolic dependence of $M_{p}$
versus $m_{\rm H}$ for three different metallicities and three different stellar
masses. \label{tab:tab1}}
\tablewidth{0pt}
\tablehead{\multicolumn9c{Coefficients of $M_{p}$ in equation (4) }}
\startdata
{} & \multicolumn3c{M = 1.5 \ms} & \multicolumn3c{M = 2.0 \ms} & \multicolumn3c{M = 3.0 \ms} \\
\hline
[Fe/H] & $a$ & $b$ & $c$ & $a$ & $b$ & $c$ & $a$ & $b$ & $c$ \\
\hline
-0.50 & -1.3300 & 1.7040 & -0.5414 & -0.5070 & 0.6680 & -0.2164 & 0.019 & 0.0426 & -0.0366 \\
-0.30 & -2.0050 & 2.4775 & -0.7594 & 0.9646 & -1.2658 & 0.4180 & -0.8647 & 1.0337  & -0.3042 \\
 0.00 & -6.6146 & 8.3383 & -2.6217 & -1.3566 & 1.7393 & -0.5527 & -0.1373 & 0.1396 & -0.0298 \\
\enddata
\end{deluxetable*}

As mentioned, the proton distribution resulting in the above layers can be computed by the
equations presented in paper II. The extensions in mass $M_p$ of the reservoirs vary in 
roughly quadratic (parabolic) ways as a function of the core mass $m_{\rm H}$  (which specifies 
the moment in the AGB evolution when a given TDU episode occurs). Regression
coefficients are in this case between 0.97 and 0.99. 

\begin{figure*}[t!!]
\begin{center}
\includegraphics[height=\textheight]{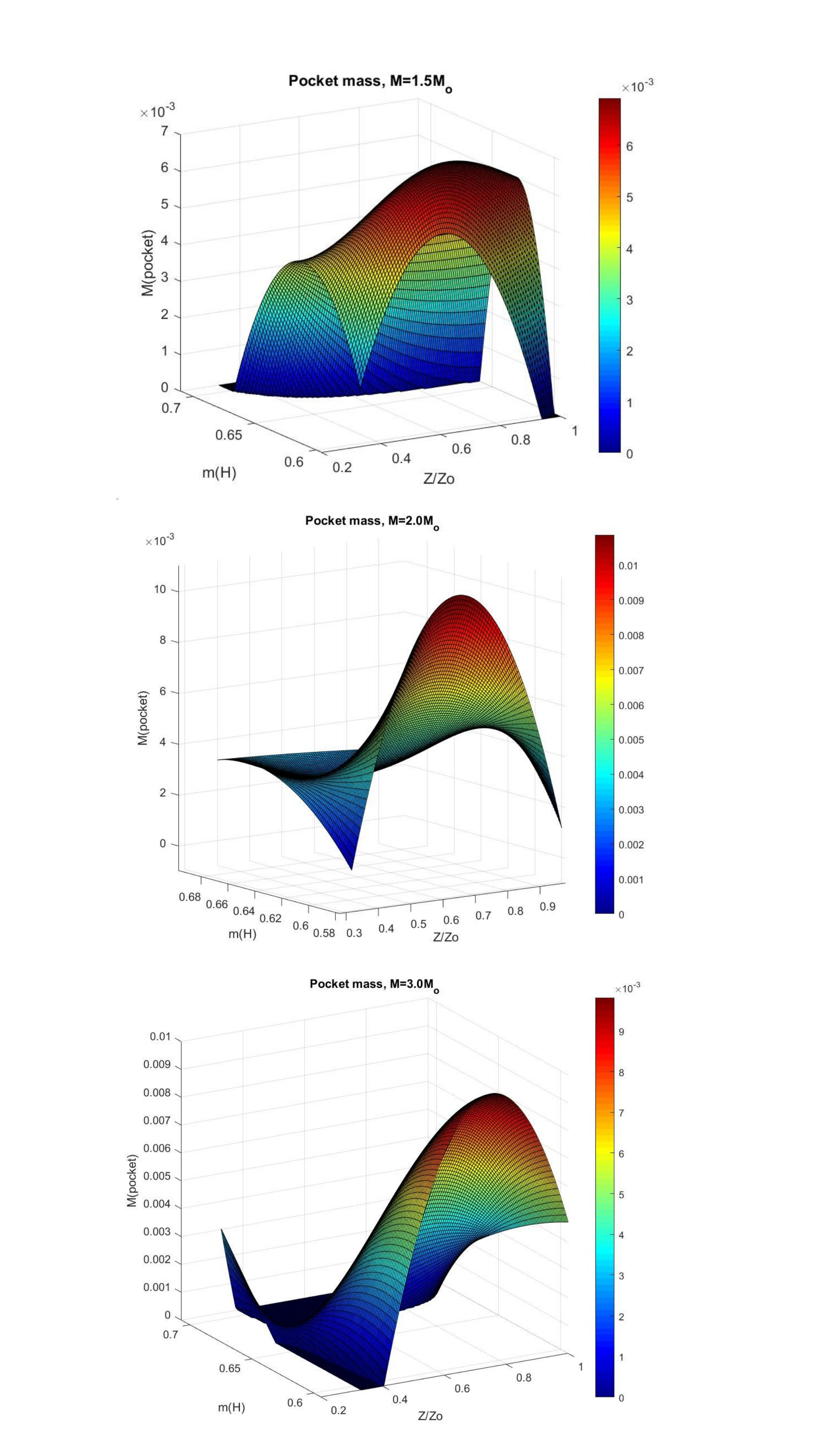}
\caption{A synthetic view of the dependence of the pocket mass $M_p$ on the 
core mass ($m_{\rm H}$) and the relative metallicity ($f$ = X(Fe)/X(Fe)$_{\odot}$) for 
the reference models, at typical Galactic disk metallicities. The plots 
are pure fits to model results and cannot be extrapolated beyond their limits. Note that the
apparent {\it spike} at the low left end in the bottom panel is real: in the FRANEC 
code, at low metallicity, a 3.0 \msb model star behaves almost as an IMS, with TDU (hence also \ctb pockets) starting at relavely high values of the core mass.\label{fig:pockmass}}
\end{center}
\end{figure*}

Indicating by $f$ the relative metallicity in a linear scale 
($f = {\rm Fe}/{\rm Fe}_{\odot} = 10^{\rm [Fe/H]}$), we can then write:
$$
M_{p} = a(f)\cdot m_{\rm H}^2 + b(f)\cdot m_{\rm H} +c(f) \eqno(4)
$$
where the three coefficients $a(f)$, $b(f)$, $c(f)$ are presented in 
Table \ref{tab:tab1}. We fitted the dependence of the coefficients on $f$
through further quadratic forms only for the sake 
of illustration, albeit there is no physics in this procedure, just 
an analytic formulation given for convenience. The dependence
on the two parameters $m_{\rm H}$ and $f = {\rm Fe}/{\rm Fe}_{\odot}$, is then
shown in Figure \ref{fig:pockmass} for the reference stellar models, over 
a range of metallicities typical of the Galactic disk

Once the extension of the \ctb reservoirs and the profile in them
of the \ctb abundance were estimated for every TDU episode of the 
mentioned stellar models, we used the NEWTON post-process code 
(see paper II) for computing the nucleosynthesis {induced by} the
combined activation of the neutron sources \ctab and \nean.

\begin{figure*}[t!!]
\begin{center}
\includegraphics[width=0.85\textwidth]{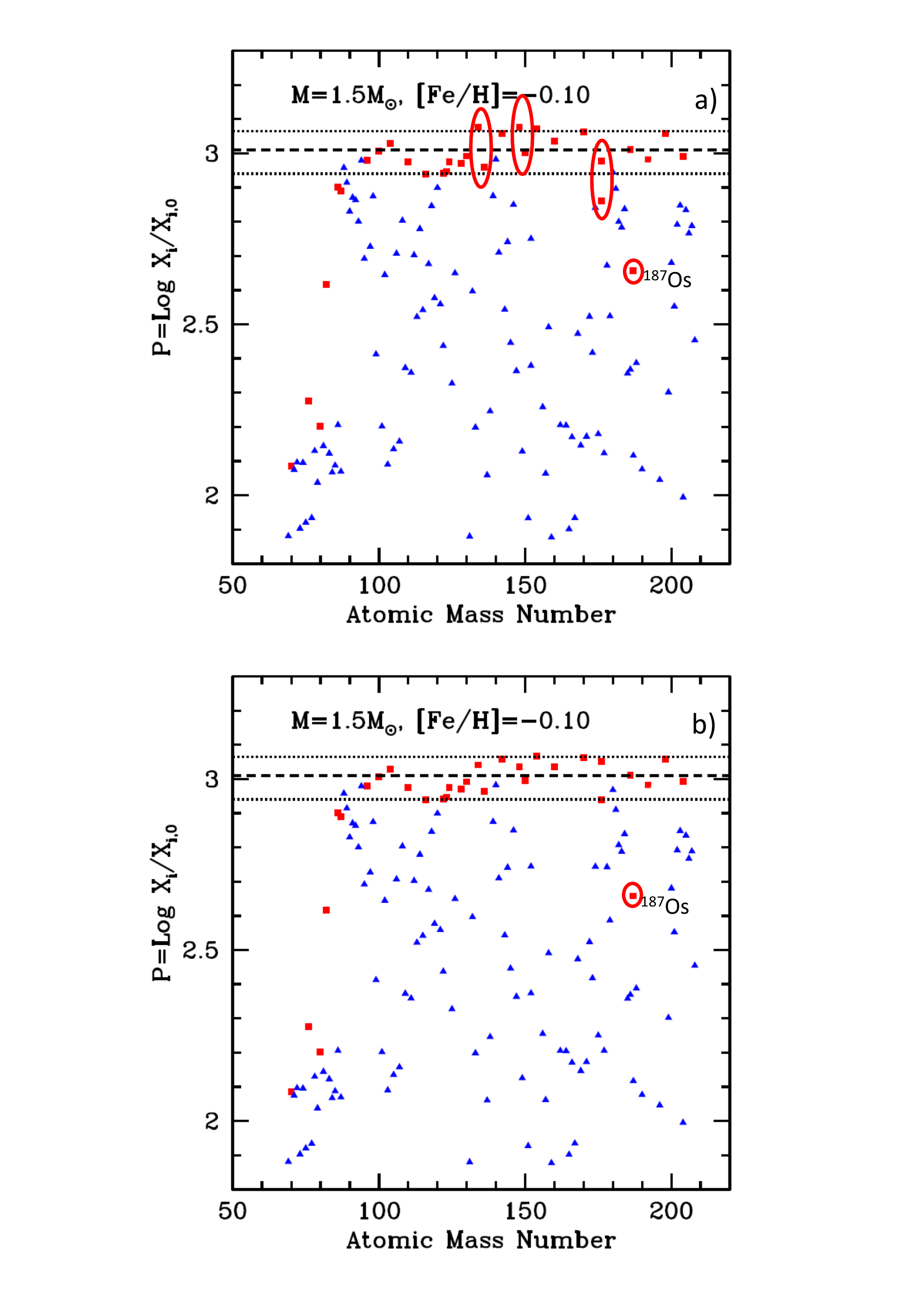}
\caption{The production factors of nuclei beyond Fe in the He-intershell layers for the model
of a 1.5 \msb star of metallicity slightly lower than solar, mimicking rather well
the solar distribution of $s$-only nuclei (red squared dots). The average overabundance for
$s$-only nuclei is slightly higher than 1000 (dashed line: a general fiducial
uncertainty of 15\% is indicated through the dotted lines). The other symbols 
(in blue) represent nuclei only partly contributed by $s$-processing. Panel a) shows the discrepancies 
found on close-by $s$-only nuclei in the preliminary computations made with the nuclear 
parameters mentioned in the text. The alternative choices we suggest permit to reduce the 
scatter considerably, as shown in panel b).   \label{fig:singlemod}}
\end{center}
\end{figure*}

As in previous issues of this series of works, the post-process code carefully
imports from the stellar model the relevant physical parameters (extension
of the intermediate convective layers at TPs, their temperature and density profiles in mass and time, the timing and extension of dredge-up phenomena, 
the mass of the envelope, gradually eroded by the growth of the core mass 
m$_{\rm H}$ and by mass loss, etc.). Reference solar abundances were taken from \citet{lod1}. The cross section 
database is from the on-line compilation KADONIS, v0.3 \citep{dil1}, with integrations from the upgrade 
KADONIS v1.0 \citep{dil2}. These extensions require some careful considerations, in particular for the stellar enhancement factors (SEFs). Indeed, in KADONIS v1.0, these corrections are accompanied with the alternative suggestions
proposed by \citet{rauscher}, called $X-factors$. These last are sometimes sharply different from the traditional SEFs used by many groups and also by us so far. The choice of which stellar corrections to apply requires therefore some
iterative checks, to be performed on preliminary computations (see later and Figure \ref{fig:singlemod}). Recent results from the n$\_$TOF collaboration \citep{ntof0,ntof1} are also included. Concerning weak interactions, their rates are
normally taken from  \citet{taka1, taka2}. For the cases in which bound-state decays are known to occur in ionized 
plasmas, the  corrections discussed in \citet{taka3} and in Table V of \citet{taka1} were applied, although these
are probably still insufficient to mimic real stellar conditions.

On this point we must remember that weak interaction rates in stars remain the largest source of uncertainty in $s$-processing. 
This is e.g. evident in the important case of the $s$-only nucleus $^{187}$Os (produced by decay of 
its parent $^{187}$Re, which in laboratory is a very-long lived isotope, with $t_{1/2}^{lab} \simeq$  40 Gyr).
The parent is known to undergo a bound-state decay in stellar conditions: due to this fact, when it is 
completely ionized it decays in few tens of years \citep{lit}. The problem is that $s$-process conditions should correspond to partial ionization, and one would need to interpolate over nine orders of magnitude. We adopted here 
the Local Thermodynamic Equilibrium approach by \citet{taka1} for the dependency of the $^{187}$Re rate on temperature 
and density, but the treatment is largely insufficient. It is moreover unknown what kind of change is induced on the
$^{187}$Re half-life by the process of astration, due to which the nucleus enters successive stellar generations, where it
is taken at high temperature, so that its decay rate certainly increases, but we do not know by how much. As a consequence, 
in our models $^{187}$Os is produced only partly (see Figure 4b and later Figure 14). We are forced, in this respect, 
to accept that the quantitative reproduction of its abundance must wait for new measurements in conditions
simulating the stellar ones.

Similar uncertainties exist for several nuclei immediately following reaction branchings in 
the $s$-process path, where weak interaction rates dominate the production. A remarkable example 
is that of the couple $^{176}$Lu-$^{176}$Hf (see Figure \ref{fig:luhf}). The first nucleus is 
a long-lived isotope in laboratory conditions \citep[with a half-life of 36 Gyr, see][]{soder}. In
stars it presents a short-lived isomeric state. Direct link of this with the ground state via dipole transitions 
is forbidden, at the temperature where \ctb burns (0.9$\cdot 10^8$ K), so that the two 
states of $^{176}$Lu effectively behave as separate nuclei. However, when in the AGB environment 
a thermal instability develops, the locally high temperature ($T \ge 3 \cdot 10^8$) can excite a number of 
overlying mediating states and the isomeric level gets thermalized. These complications, studied in 
detail by  \citet{klay1, klay2, k6}, sum to the fact that the half-life above about 20$-$22 keV 
becomes very short and dependent on temperature, so that $^{176}$Lu
actually behaves as a thermometer \citep{klay2}. In our results, $^{176}$Lu and its daughter  
$^{176}$Hf are formally inside a $reasonable$ general error bar of $\simeq$ 15\% (see Figures \ref{fig:singlemod} and \ref{fig:bestdist}) but at the extremes of it; any further improvement must wait for better nuclear inputs. Another important case of weak interaction effects concerns $^{134}$Cs, whose 
$\beta^-$ decay to $^{134}$Ba is crucial in fixing the abundance ratio of 
the two $s$-only isotopes $^{134}$Ba and $^{136}$Ba. The value taken from \citet{taka1} 
and its temperature dependence is certainly very uncertain \citep[see discussion in][]{k+90}.  
Again, we must be content that, despite this, 
the model abundances of $^{134}$Ba and $^{136}$Ba lay within the general fiducial 
error bar; improving on this would require dedicated experimental data. In fact, some of the relevant radioactive 
nuclei along the $s$-path suitable to be affected by these specific uncertainties are now in the 
program of the new experiment PANDORA \citep{pan}, which will start in 2022 for measuring decay rates in 
ionization conditions as similar as possible to the stellar ones.

As already suggested in paper II, $s$-processing in the Galaxy has the remarkable property 
that one can identify a specific model (characterized by a low initial mass and a metallicity 
typical of the Galactic disk within 2 $-$ 3 Gyr before the Sun is formed), which roughly simulates the 
solar distribution. Figure \ref{fig:singlemod} shows one such case, where indeed the abundances of $s$-only 
nuclei (indicated by heavy squared dots) have similar production factors, roughly
averaging at about 1000. This model grossly represents a sort of average of what can be 
more properly obtained by the chemical evolution of the Galaxy, but this last is of course needed
to account for the different effects of stellar temperatures in differently massive stars. The 
$average$ model is however a suitable preliminary calculation to be performed, on which 
to test tentatively the uncertain corrections to cross sections mentioned above.

The top panel of Figure \ref{fig:singlemod} (panel a) shows results computed using the corrective 
$X-factors$ to cross sections from \citet{rauscher}. It is evident that, although the distribution 
for $s$-only nuclei (red dots) is rather flat (as it should be for producing them in solar proportions),
there exist cases in which adjacent $s$-only nuclei show considerably discrepant abundances (see
the red evidencing marks). If one normalizes the average to 1.0, then these discrepancies become evident
in particular for $^{134}$Ba (1.21) and $^{136}$Ba (0.89); then for $^{148}$Sm (1.20) and $^{150}$Sm (1.01) 
and for $^{176}$Lu (0.95) and $^{176}$Hf (0.73). All these are complex cases, affected also by large 
uncertainties on $\beta^-$-decay rates (for  $^{134}$Cs, $^{149}$Sm, $^{176}$Lu$^g$ and its isomer 
$^{176}$Lu$^m$). For all the nuclei considered, however, a
significant worsening of the distribution is induced by the application of the mentioned $X-factors$, which 
sometimes imply corrections opposite to the traditional SEFs. Although the uncertainties do not allow 
us to reject the $X-factor$ corrections in general, we made an alternative computation by changing them with 
the usual SEFs to verify which data set was more suitable for obtaining a solar-like distribution. The result 
of this test is presented in Figure \ref{fig:singlemod}, panel b), where it is shown that, for the three 
couples indicated before, some improvements are immediately obtained. The new ratios found in the $average$ 
model are: 1.09 ($^{134}$Ba),
0.92 ($^{136}$Ba), 1.08 ($^{148}$Sm), 0.99 ($^{150}$Sm), 1.11 ($^{176}$Lu), 0.88 ($^{176}$Hf). On this basis, 
we decided to adopt, in the rest of this paper, the common SEFs to cross sections, postponing a more detailed 
check of the corrections by \citet{rauscher} to a separate work.

\begin{figure}[t!!]
\begin{center}
\includegraphics[width=0.85\columnwidth]{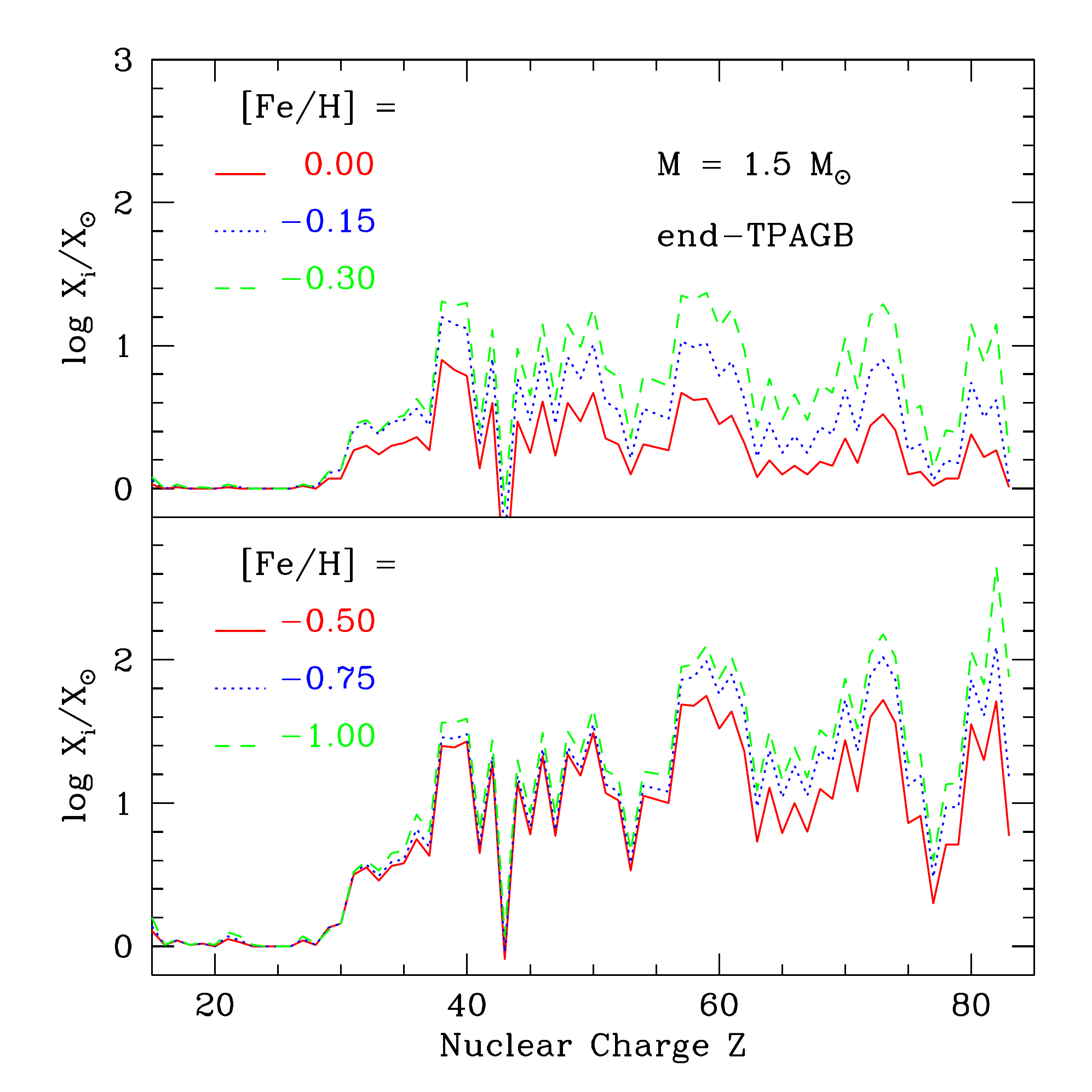}
\caption{Abundances in the envelope at the last TDU episode computed, for stars of 
initial mass M = 1.5 \ms, at the indicated metallicities. \label{fig:lastm1p5}}
\end{center}
\end{figure}

\begin{figure}[t!!]
\begin{center}
\includegraphics[width=0.85\columnwidth]{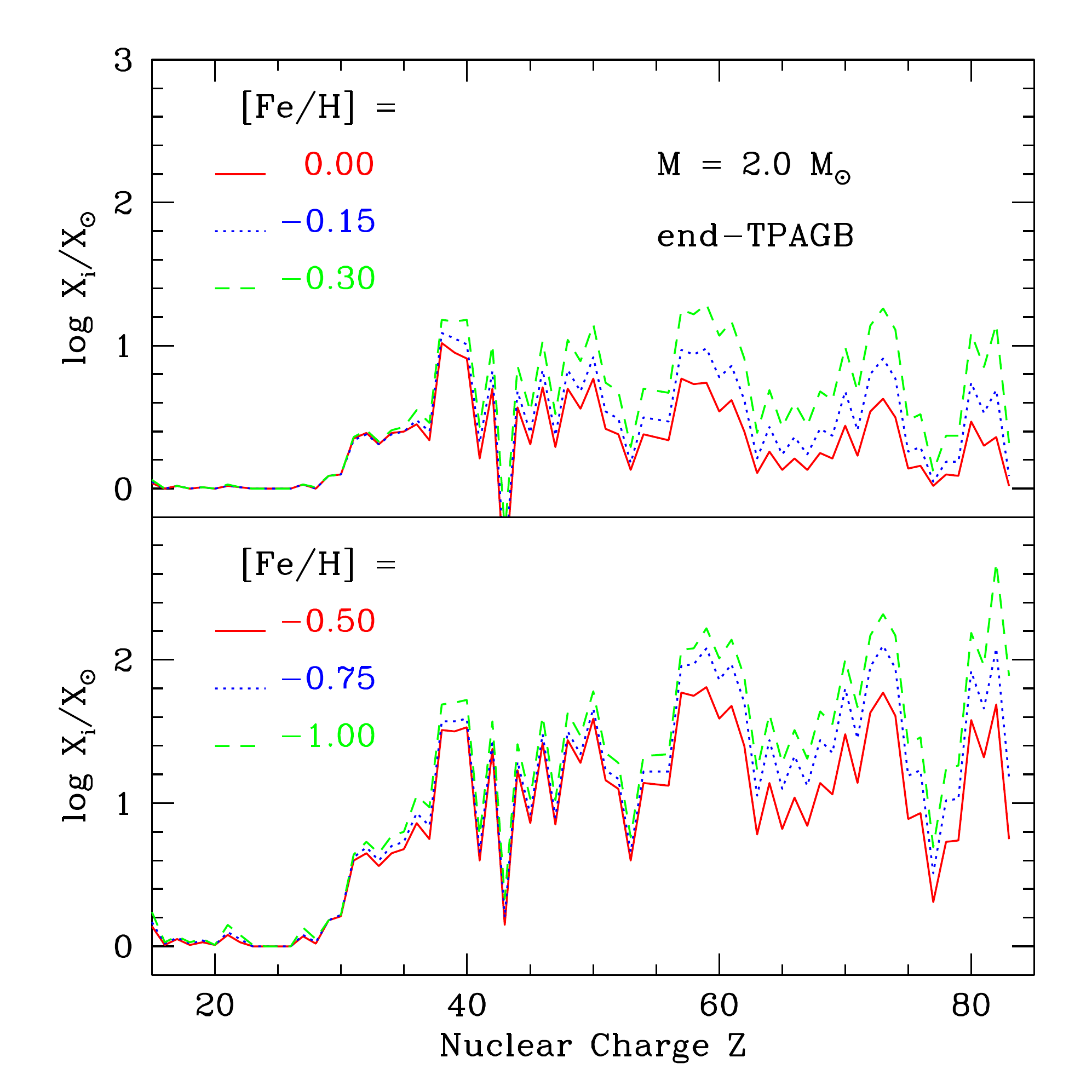}
\caption{Same as in Figure \ref{fig:lastm1p5}, for models of 2.0 \ms. \label{fig:lastm2}}
\end{center}
\end{figure}

\begin{figure}[t!!]
\begin{center}
\includegraphics[width=0.85\columnwidth]{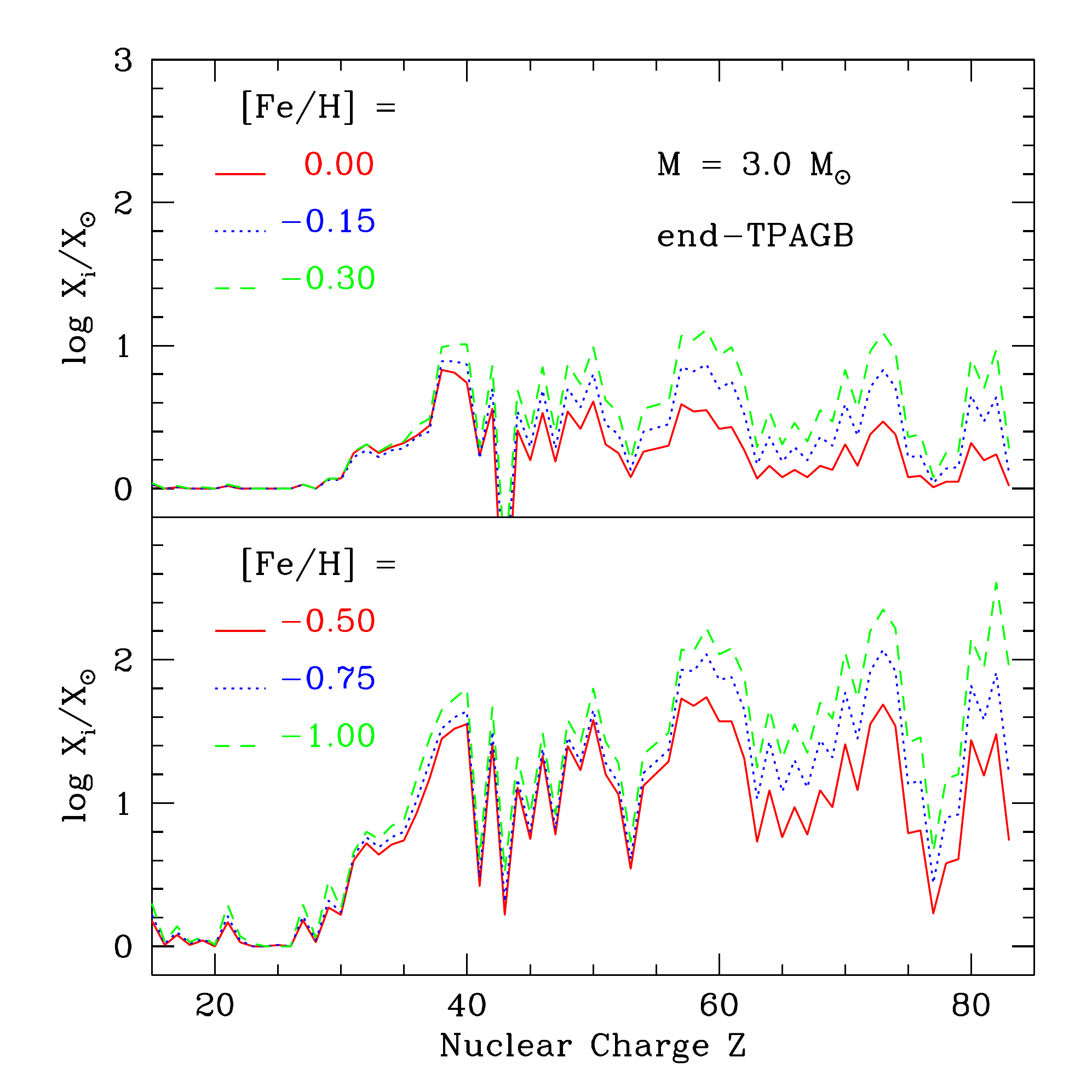}
\caption{Same as in Figure \ref{fig:lastm1p5}, for models of 3.0 \ms. \label{fig:lastm3}}
\end{center}
\end{figure}

With the choices thus made, examples of abundance distributions in the envelopes at the end of the TP-AGB 
stage are shown in Figures \ref{fig:lastm1p5}, \ref{fig:lastm2} 
and \ref{fig:lastm3} for various stellar masses and metallicities. The number of TDU episodes found in the cases
shown by the figures is reported in Table \ref{tab:tab2}.

\begin{deluxetable*}{c|ccc|ccc|ccc}
\tablenum{2}
\tablecaption{General characteristics of the TDU phases where the conditions for forming \ctb pockets are found, according to the criteria exposed in the text. \label{tab:tab2}}
\tablewidth{0pt}
\tablehead{\multicolumn9c{Number of TDU episodes and their Maximum/minimum extension in mass}}
\startdata
{} & \multicolumn3c{M = 1.5 \ms} & \multicolumn3c{M = 2.0 \ms} & \multicolumn3c{M = 3.0 \ms} \\
\hline
[Fe/H] & $N$ & $\Delta M_{min}$ & $\Delta M_{Max}$ & $N$ & $\Delta M_{min}$ & $\Delta M_{Max}$ & $N$ & $\Delta M_{min}$ & $\Delta M_{Max}$ \\
\hline
-0.50 & 11 & 3.6$\cdot$10$^{-4}$ & 1.5$\cdot$10$^{-3}$ & 13 & 1.5$\cdot$10$^{-3}$ & 3.6$\cdot$10$^{-3}$ & 11 & 2.4$\cdot$10$^{-4}$ & 3.7$\cdot$10$^{-3}$ \\
-0.30 & 10 & 7.2$\cdot$10$^{-4}$ & 1.5$\cdot$10$^{-3}$ & 13 & 2.2$\cdot$10$^{-5}$ & 1.7$\cdot$10$^{-3}$ & 13 & 1.0$\cdot$10$^{-4}$ & 2.2$\cdot$10$^{-3}$  \\
 0.00 & 11 & 1.5$\cdot$10$^{-4}$ & 7.1$\cdot$10$^{-4}$ & 12 & 3.8$\cdot$10$^{-4}$ & 1.6$\cdot$10$^{-3}$ & 17 & 1.0$\cdot$10$^{-4}$ & 2.1$\cdot$10$^{-3}$  \\
\enddata
\end{deluxetable*}

The distributions presented in Figures \ref{fig:lastm1p5} to \ref{fig:lastm3} illustrate the increase in 
the abundances of $n$-capture elements expected for metal-poor AGB stars, 
a trend that had been previously inferred from parametric models 
\citep[see e.g.][]{gal1} and in general permits to account for the
observations of $s$-elements in AGB stars at different metallicity.
This appears now to remain true even in computations where the \ctb pocket 
is generated by a physical model not explicitly related to
metallicity. This is so because the complex dependence of the pocket 
masses on the stellar parameters shown in Figure \ref{fig:pockmass} is not
sufficient to erase the signatures impressed in the $s$-process distribution
by the fact that, for a {\it primary-like} neutron source, the neutron 
exposure grows for decreasing abundances of the seeds (mainly Fe) 
by which the neutrons are captured.

\begin{figure}[t!!]
\begin{center}
\includegraphics[width=0.85\columnwidth]{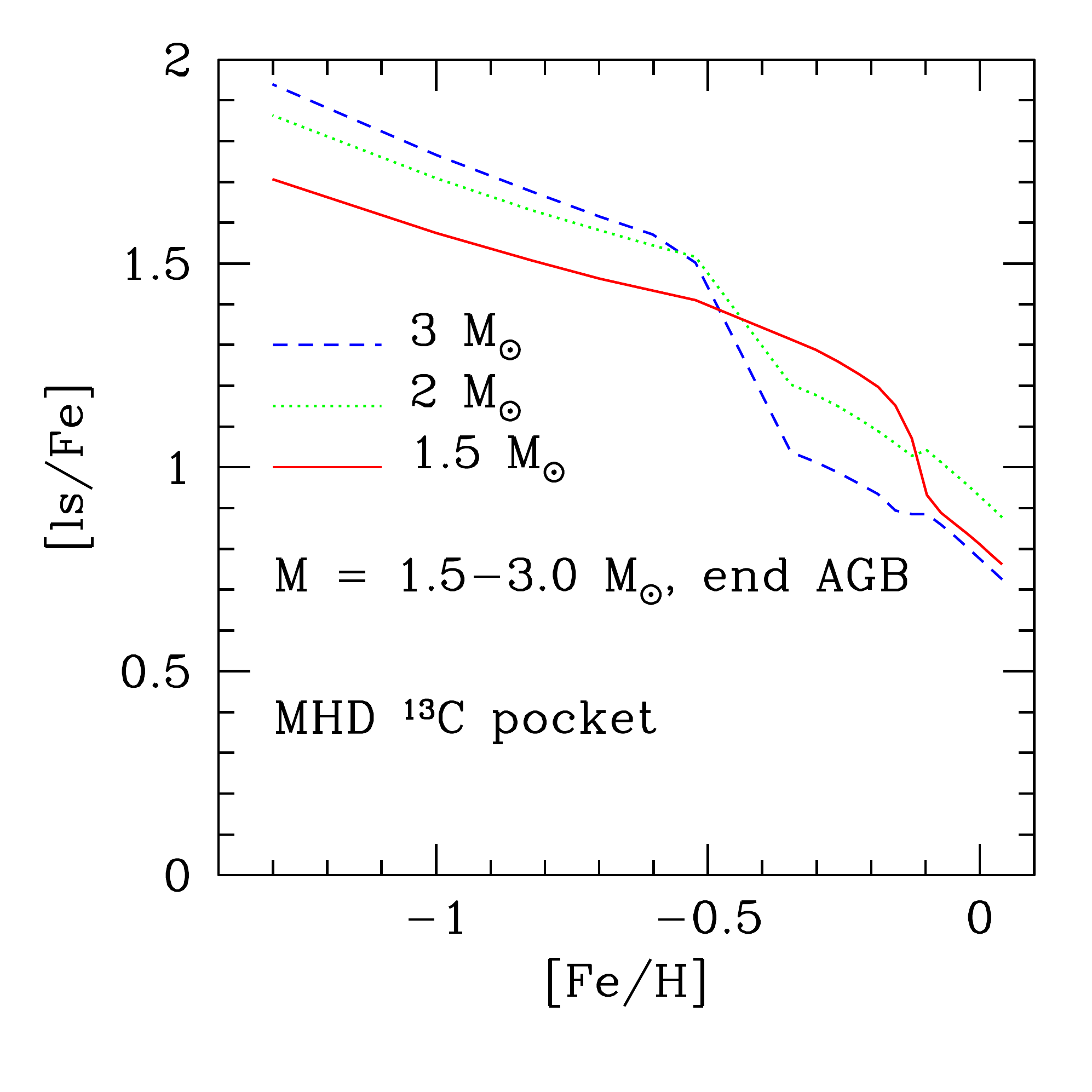}
\caption{Production factors for elements at the abundance peak near the
magic neutron number $N=50$, as summarized by the $[ls/{\rm Fe}]$ indicator, for  
stellar masses $M =$ 1.5, 2.0 and 3.0 \msb at various metallicities. \label{fig:lsfe}}
\end{center}
\end{figure}

\begin{figure}[t!!]
\begin{center}
\includegraphics[width=0.85\columnwidth]{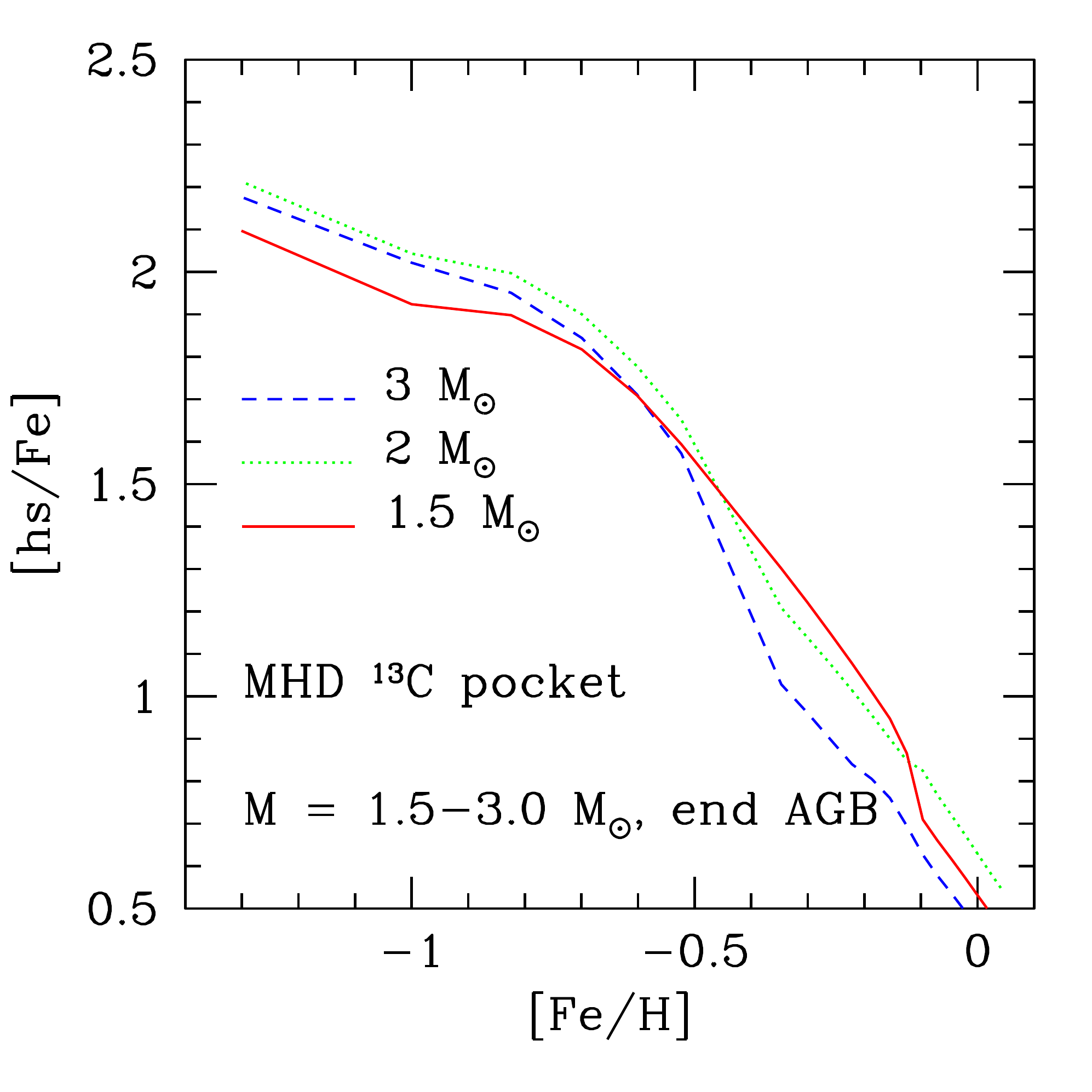}
\caption{Production factors for elements at the abundance peak near the
magic neutron number $N=82$, as summarized by the $[hs/{\rm Fe}]$ indicator, for  
stellar masses $M =$ 1.5, 2.0 and 3.0 \msb at various metallicities. \label{fig:hsfe}}
\end{center}
\end{figure}

\begin{figure}[t!!]
\begin{center}
\includegraphics[width=0.85\columnwidth]{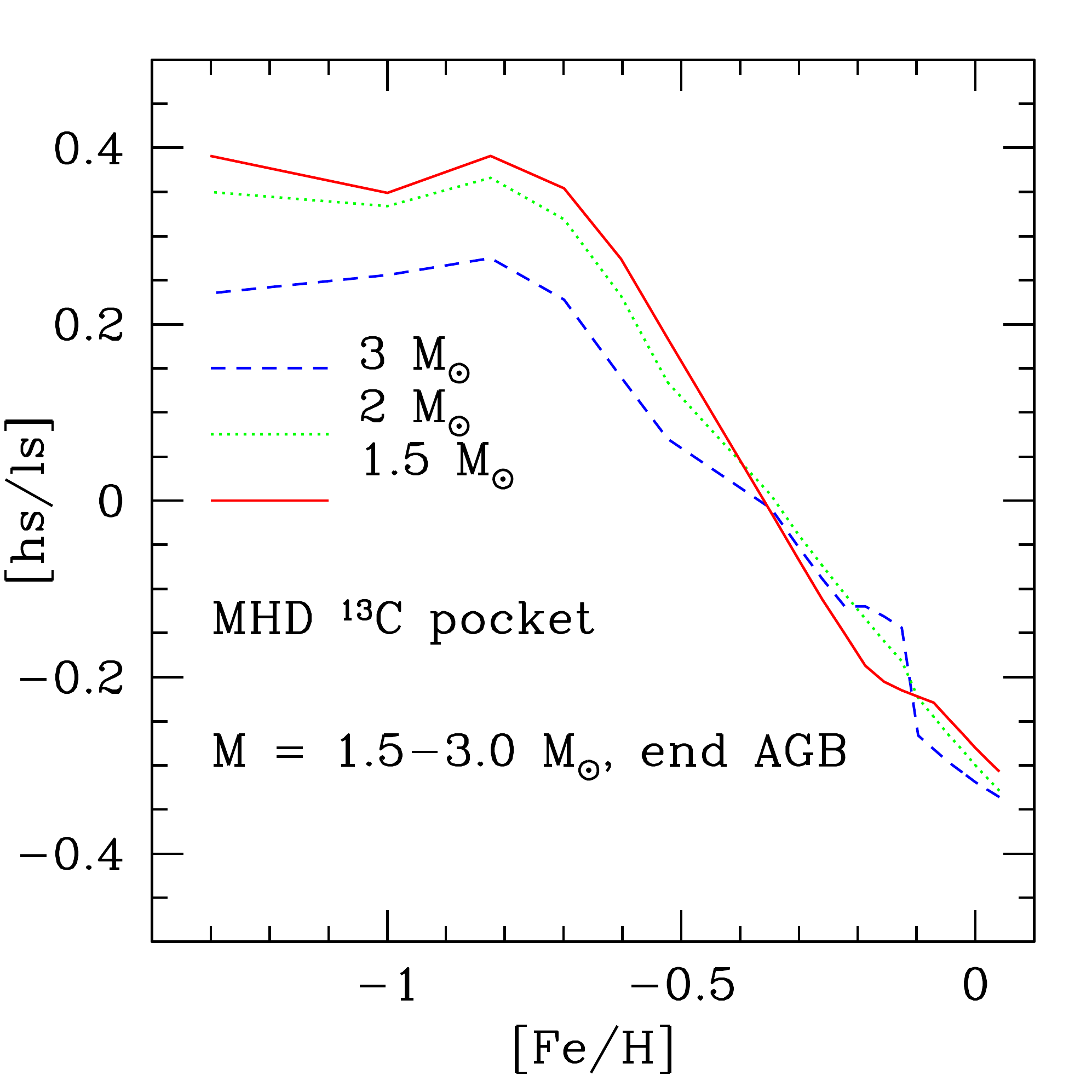}
\caption{The logarithmic ratios of abundances at the two main $s$-process peaks
for $N=50$ and $N=82$, as summarized by the $[hs/ls]$ indicator, for the same cases 
shown in the previous figures. \label{fig:hsls}}
\end{center}
\end{figure}

\begin{figure}[t!!]
\begin{center}
\includegraphics[width=0.85\columnwidth]{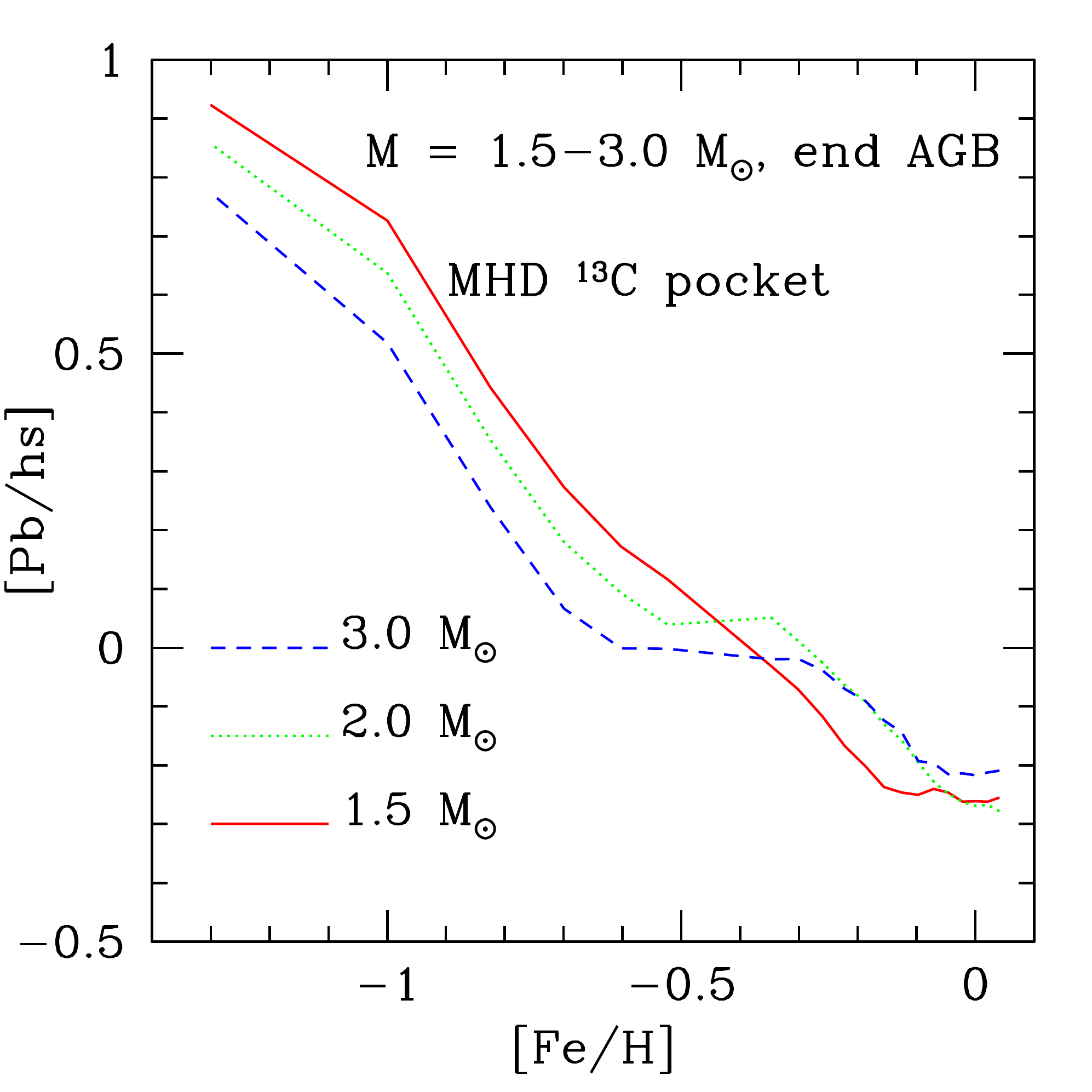}
\caption{Relative production factors for Pb and for elements at the abundance peak near the
magic neutron number $N=82$, as summarized by the $[{\rm Pb}/hs]$ indicator, for  
stellar masses $M=$ 1.5, 2.0 and 3.0 \msb at various metallicities. \label{fig:pbhsagb}}
\end{center}
\end{figure}

Since the suggestions by \citet{lb1,lb2} it is common to synthetically 
represent the $s$-process distribution in stars with the two indices $[ls/{\rm Fe}]$ 
and $[hs/{\rm Fe}]$, where {\it ls} stands for {\it light s-elements} and {\it hs} stands for 
{\it heavy s-elements}. They are built with the logarithmic abundances at the 
first and second $s$-process peak, near the neutron magic numbers $N=50$ and $N=82$. 
In particular, we follow suggestions by \citet{lb2} in assuming [$hs$/{\rm Fe}] =
0.25$\cdot$([Ba/Fe]+[La/Fe]+[Nd/Fe]+[Sm/Fe]), and by \citet{b+01} in assuming 
[$ls$/{\rm Fe}]=0.5$\cdot$ ([Y/Fe]+[Zr/Fe]). Their difference, $[hs/ls] = [hs/{\rm Fe}] - 
[ls/{\rm Fe}]$, 
is an effective indicator of the neutron exposure, with low neutron fluences
feeding primarily the $[ls]$ abundances and high fluences making the $[hs]$ indicator
to prevail. In figures \ref{fig:lsfe}, \ref{fig:hsfe} and \ref{fig:hsls} the 
behavior of neutron-capture elements as a function of metallicity in our 
models is illustrated through the indices thus defined. As clarified many years
ago \citep{gal1}, the heaviest $s$-process isotopes (the so-called $strong$ component
of the $s$-process), and in particular $^{208}$Pb,
grow with metallicity with a trend that is steeper than for the $[hs]$ nuclei, 
the heaviest isotopes being sited 
in correspondence of another magic number of neutrons ($N = 128$). This property is preserved 
in the present scenario. Figure \ref{fig:pbhsagb} shows the further increase at low
metallicity of lead with respect to the $[hs]$ nuclei, these last representing the $N=82$ neutron 
magic number. We notice that the trend of $[{\rm Pb}/hs]$ shown in the figure is in nice agreement
with what is obtained by a few current models with parametric extra-mixing, like e.g. those of the
STAREVOL and FRUITY collaborations. For this, see in particular \citet{desmedt4} and Figure 14 
in \citet{desmedt2}.

\section{Reproducing constraints from the Sun and recent stellar populations}

\begin{figure}[t!!]
\begin{center}
\includegraphics[width=0.9\columnwidth]{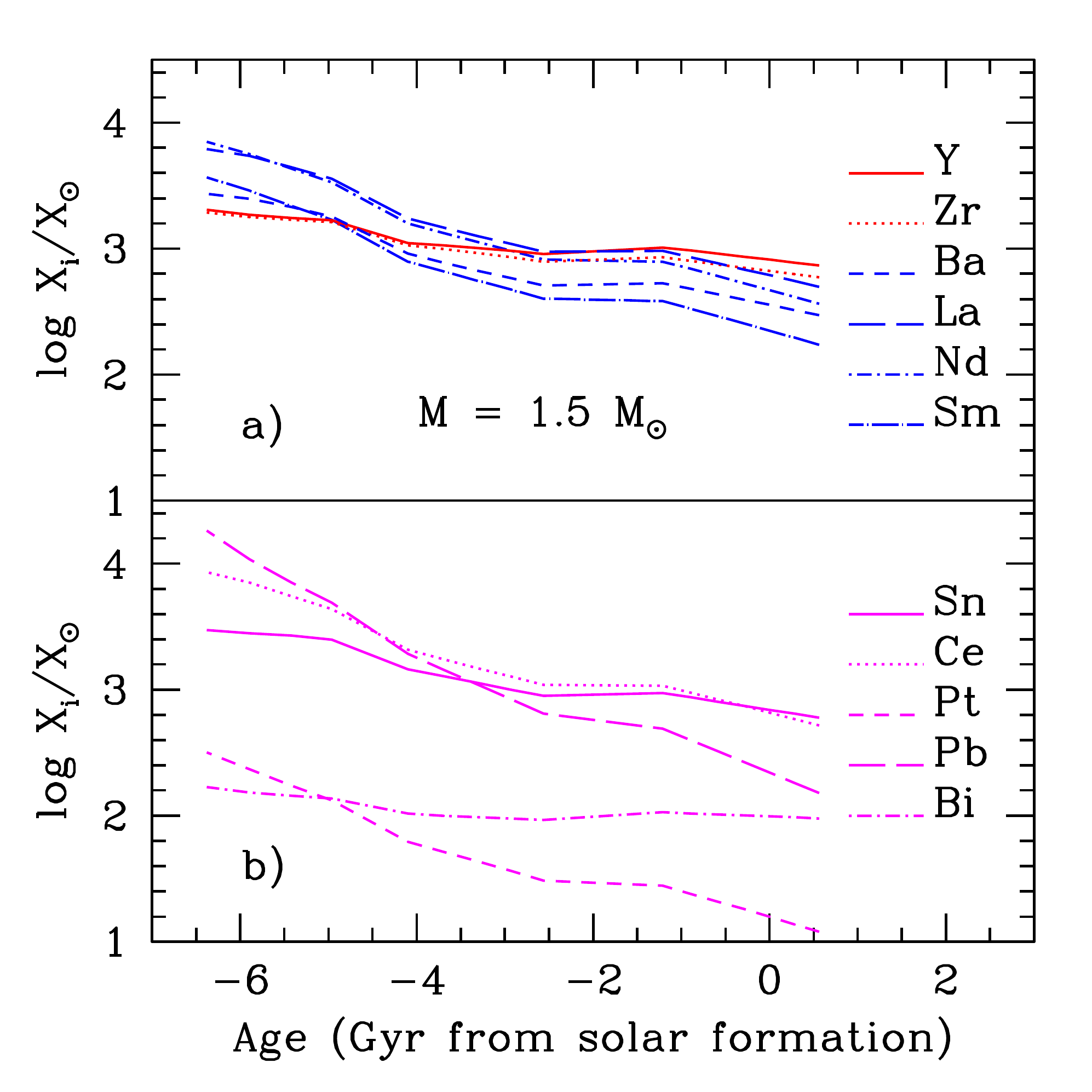}
\caption{Production factors of representative neutron-rich 
elements in stellar models of 1.5 \ms, for varying metallicity and Galactic
age, using the Age-Metallicity relation mentioned in the text. The elements at the
first (red) and second (blue) $s$-process peak (panel a) remain remarkably similar 
for a rather long time in the Galactic evolution that preceded solar formation 
(roughly between -4 and -1.5 Gyr). This is not true for the other elements (panel b, 
magenta lines): in particular, Pb has the peculiar trend of increasing steadily for
decreasing metallicity. \label{fig:elage1p5}}
\end{center}
\end{figure}

\begin{figure}[t!!]
\begin{center}
\includegraphics[width=0.9\columnwidth]{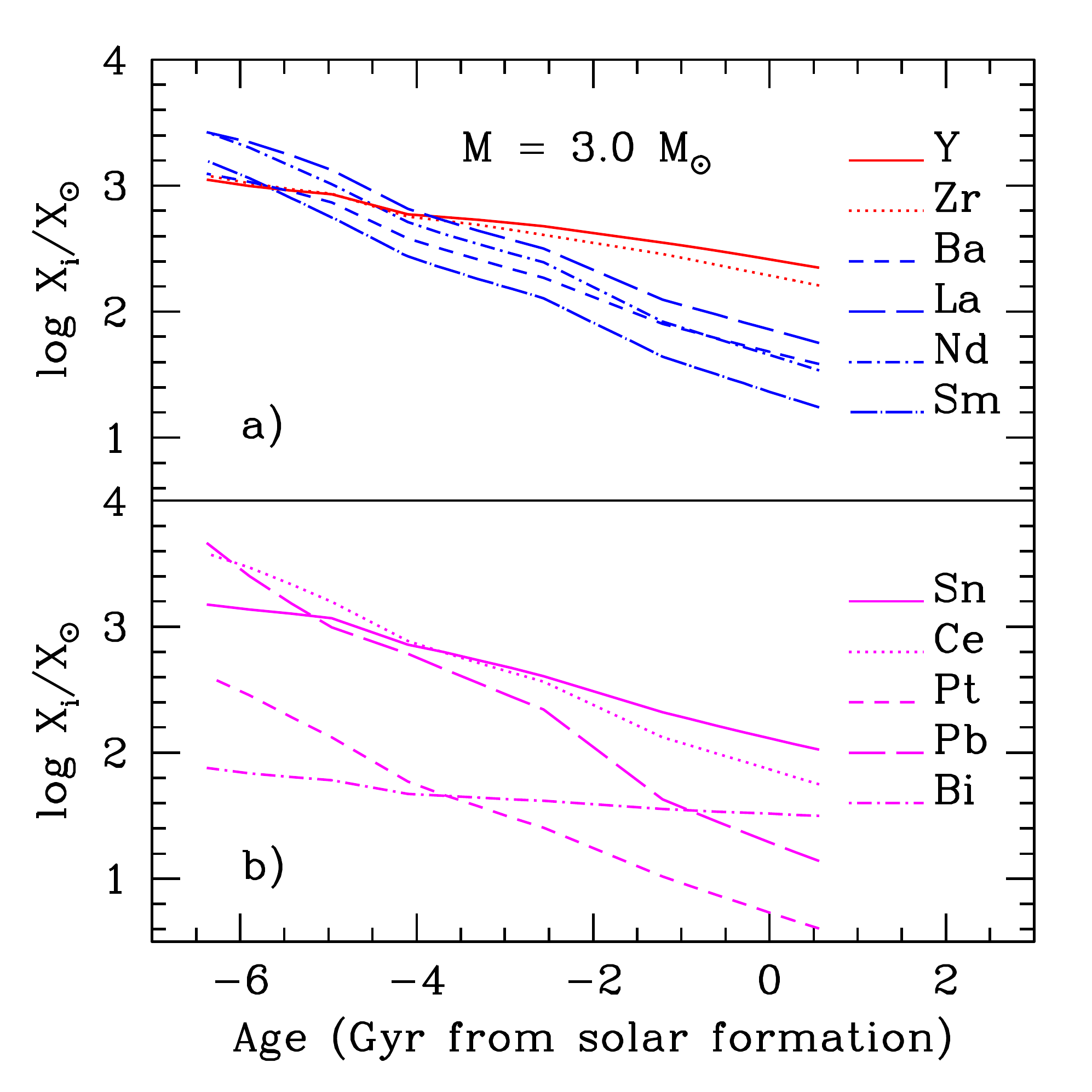}
\caption{Same as in Figure \ref{fig:elage1p5}, for models of 3.0 \ms. Here the stationary behavior shown in panel a) of the previous figure is not present, if not for a shorter time interval at earlier ages. \label{fig:elage3}}
\end{center}
\end{figure}

The effects of an efficiency in $s$-processing that increases toward lower 
metallicities are then mediated by the rate at which stars form as a function
of the metal content of the Galaxy and of its age, these two parameters being 
linked by an extremely non-linear and probably non-unique \citep{casali} relation. 
In order to mimic abundances of the solar neighborhoods, we used an Age-Metallicity 
relation that was shown to be valid for that region \citep{mai1,mai2} to 
switch between the original metal content of a stellar model and the Galactic age 
to which its formation roughly corresponds. The results are shown in Figures 
\ref{fig:elage1p5} and \ref{fig:elage3} for the extreme cases of stellar masses 
M=1.5 \msb and M=3.0 \ms.

As Figure \ref{fig:elage1p5} illustrates (see its panel a), for low stellar masses the production
factors of elements near the two main $s$-process abundance peaks remain very similar, confined in a small range of less than 0.3 dex, for quite a long period in the Galactic disk before the solar metallicity [Fe/H] $= 0$ is reached (a few Gyr). This is not the case for larger masses (see Figure \ref{fig:elage3}), which however have a lower weight
in the IMF. As discussed in \citet{mai2}, this condition is essential to permit a growth of $s$-element abundances that continues after the epoch of the solar formation maintaining a roughly constant $[hs/ls]$ ratio, as observed in Young Open Clusters. 
We are therefore confident that our scenario fulfills this basic constraint of Galactic chemical evolution, previously
met only by varying parametrically the amount of \ctb burnt and its distribution.

\begin{figure}[t!!]
\begin{center}
\includegraphics[width=0.85\columnwidth]{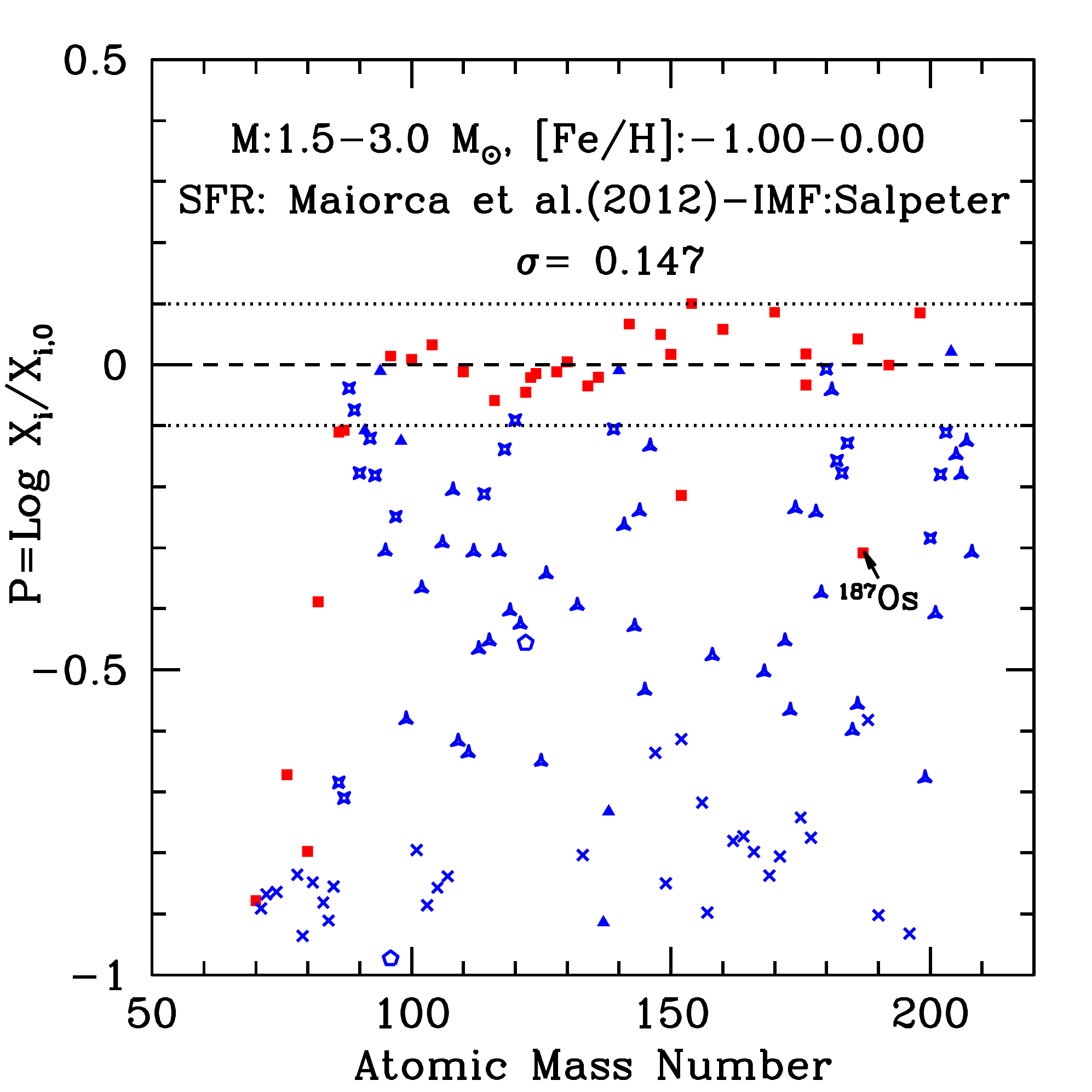}
\caption{Simulation of the chemical admixture operated by Galactic evolution,
obtained by weighting our model abundances over a Salpeter IMF and a SFR taken 
from \citet{mai2}. Symbols and colors have the same meaning as in Figure \ref{fig:singlemod} \label{fig:bestdist}}
\end{center}
\end{figure}

\begin{figure}[t!!]
\includegraphics[width=1.1\columnwidth]{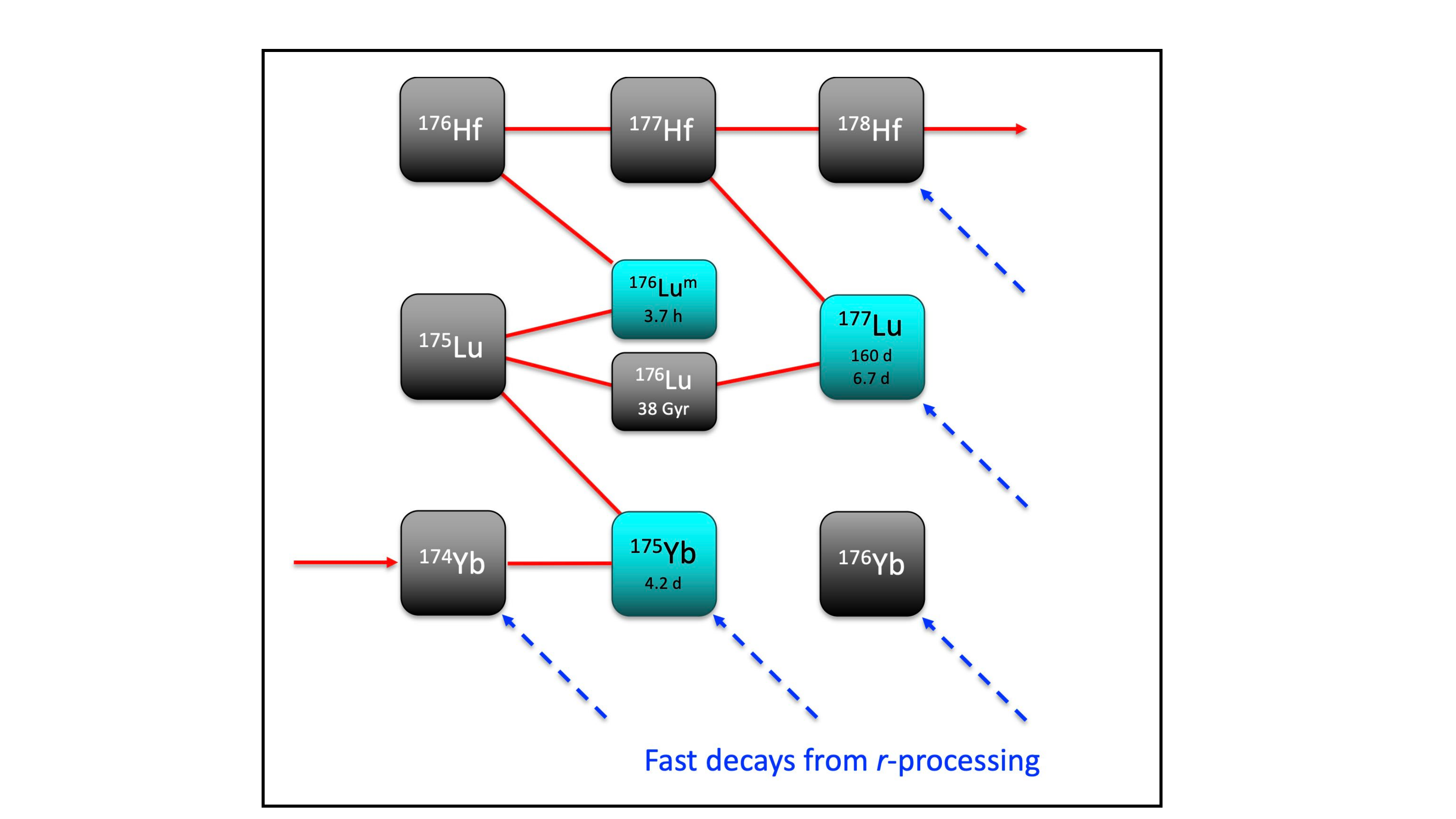}
\caption{The neutron-capture path around $^{176}$Lu and $^{176}$Hf in the chart of the nuclei, when the isomeric state and the ground state of $^{176}$Lu are not thermalized. See text for comments. \label{fig:luhf}}
\end{figure}

We have for the moment simulated such a chemical evolution by weighting the stellar 
production factors of our models for Galactic disk metallicities. We considered for this 
the range -1.0 $\lesssim $[Fe/H]$ \lesssim$ 0.0, i.e. excluding the most metal poor 
and the the most metal-rich (super-solar) models. We adopted a Salpeter initial mass 
function (IMF) and the mentioned history of the star formation rate (SFR) from 
\citet{mai2}. The result, once normalized to 1 for the average production factor of 
$s$-only nuclei, is presented in Table \ref{tab:tab3} and in Figure \ref{fig:bestdist}. 
This last shows as the solar abundances of $s$-only nuclei are reproduced at 
a sufficient level of accuracy. There is in fact a slight asymmetry, a sort of
minor deficit of $s$-only nuclei below A $\simeq$ 130 with respect to the average distribution, 
as sometimes found in the past. It was ascribed to an unknown {\it primary} nucleosynthesis
process integrating neutron captures \citep{claudia}. However, in our results the asymmetry remains 
within the relatively small limits set by abundance and nuclear uncertainties, so that no 
real conclusion in favor of possible different nuclear processes can be drawn from our findings,
confirming previous indications by \citet{mai1, cris2b, nikos}

\begin{deluxetable*}{ccc|ccc|ccc|ccc}
\tablenum{3}
\tablecaption{Percentages from the $s$-Process Main Component\label{tab:tab3}}
\tablewidth{0pt}
\tablehead{}
\startdata
\hline
{A} & Elem. & Perc. MC & A & Elem. & Perc. MC & A & Elem. & Perc. MC &A & Elem. & Perc. MC \\
\hline
    58 & Fe & 0.031 &  63 & Cu & 0.014 &  64 & Ni & 0.047 &  64 & Zn & 0.006 \\
    65 & Cu & 0.038 &  66 & Zn & 0.026 &  67 & Zn & 0.037 &  68 & Zn & 0.059 \\
    69 & Ga & 0.083 &  70 & Ge & 0.132 &  70 & Zn & 0.005 &  71 & Ga & 0.129 \\
    72 & Ge & 0.135 &  73 & Ge & 0.087 &  74 & Ge & 0.137 &  75 & As & 0.088 \\
    76 & Se & 0.213 &  76 & Ge & 0.001 &  77 & Se & 0.092 &  78 & Se & 0.145 \\
    79 & Br & 0.116 &  80 & Kr & 0.159 &  81 & Br & 0.142 &  82 & Kr & 0.409 \\
    82 & Se & 0.001 &  83 & Kr & 0.131 &  84 & Kr & 0.123 &  85 & Rb & 0.140 \\
    86 & Kr & 0.207 &  86 & Sr & 0.776 &  87 & Rb & 0.195 &  87 & Sr & 0.781 \\
    88 & Sr & 0.915 &  89 & Y  & 0.843 &  90 & Zr & 0.664 &  91 & Zr & 0.780 \\
    92 & Zr & 0.757 &  93 & Nb & 0.660 &  94 & Zr & 0.976 &  95 & Mo & 0.495 \\
    96 & Mo & 1.033 &  96 & Zr & 0.106 &  97 & Mo & 0.564 &  98 & Mo & 0.750 \\
    99 & Ru & 0.262 & 100 & Ru & 1.021 & 101 & Ru & 0.160 & 102 & Ru & 0.430 \\
   103 & Rh & 0.130 & 104 & Pd & 1.078 & 104 & Ru & 0.012 & 105 & Pd & 0.139 \\
   106 & Pd & 0.510 & 107 & Ag & 0.015 & 108 & Pd & 0.622 & 109 & Ag & 0.241 \\
   110 & Cd & 0.975 & 110 & Pd & 0.014 & 111 & Cd & 0.231 & 112 & Cd & 0.493 \\
   113 & Cd & 0.342 & 114 & Cd & 0.614 & 115 & In & 0.352 & 116 & Sn & 0.874 \\
   116 & Cd & 0.059 & 117 & Sn & 0.494 & 118 & Sn & 0.727 & 119 & Sn & 0.395 \\
   120 & Sn & 0.811 & 121 & Sb & 0.376 & 122 & Te & 0.901 & 122 & Sn & 0.350 \\
   123 & Te & 0.953 & 123 & Sb & 0.042 & 124 & Sn & 0.002 & 124 & Te & 0.967 \\
   125 & Te & 0.224 & 126 & Te & 0.453 & 127 & I  & 0.048 & 128 & Xe & 0.974 \\
   128 & Te & 0.022 & 129 & Xe & 0.038 & 130 & Xe & 1.011 & 130 & Te & 0.001 \\
   131 & Xe & 0.078 & 132 & Xe & 0.403 & 133 & Cs & 0.157 & 134 & Ba & 0.923 \\
   134 & Xe & 0.024 & 135 & Ba & 0.055 & 136 & Ba & 0.955 & 136 & Xe & 0.001 \\
   137 & Ba & 0.122 & 138 & Ba & 0.185 & 139 & La & 0.784 & 140 & Ce & 0.979 \\
   141 & Pr & 0.545 & 142 & Nd & 1.166 & 142 & Ce & 0.084 & 143 & Nd & 0.372 \\
   144 & Nd & 0.575 & 145 & Nd & 0.292 & 146 & Nd & 0.734 & 147 & Sm & 0.231 \\
   148 & Sm & 1.121 & 148 & Nd & 0.093 & 149 & Sm & 0.141 & 150 & Sm & 1.039 \\
   150 & Nd & 0.000 & 151 & Eu & 0.090 & 152 & Gd & 0.611 & 152 & Sm & 0.243 \\
   153 & Eu & 0.053 & 154 & Gd & 1.250 & 154 & Sm & 0.005 & 155 & Gd & 0.064 \\
   156 & Gd & 0.191 & 157 & Gd & 0.127 & 158 & Gd & 0.033 & 159 & Tb & 0.078 \\
   160 & Dy & 1.140 & 160 & Gd & 0.006 & 161 & Dy & 0.059 & 162 & Dy & 0.166 \\
   163 & Dy & 0.036 & 164 & Dy & 0.169 & 165 & Ho & 0.085 & 166 & Er & 0.159 \\
   167 & Er & 0.095 & 168 & Er & 0.313 & 169 & Tm & 0.145 & 170 & Yb & 1.210 \\
   170 & Er & 0.044 & 171 & Yb & 0.156 & 172 & Yb & 0.352 & 173 & Yb & 0.271 \\
   174 & Yb & 0.581 & 175 & Lu & 0.181 & 176 & Lu & 1.219 & 176 & Hf & 0.930 \\
   176 & Yb & 0.040 & 177 & Hf & 0.167 & 178 & Hf & 0.572 & 179 & Hf & 0.422 \\
   180 & Hf & 0.983 & 181 & Ta & 0.907 & 182 & W  & 0.696 & 183 & W  & 0.665 \\
   184 & W  & 0.744 & 185 & Re & 0.251 & 186 & Os & 1.102 & 186 & W  & 0.277 \\
   187 & Re & 0.033 & 187 & Os & 0.492 & 188 & Os & 0.262 & 189 & Os & 0.042 \\
   190 & Os & 0.125 & 191 & Ir & 0.019 & 192 & Os & 0.009 & 192 & Pt & 0.999 \\
   193 & Ir & 0.013 & 194 & Pt & 0.049 & 195 & Pt & 0.020 & 196 & Pt & 0.117 \\ 
   197 & Au & 0.060 & 198 & Hg & 1.210 & 198 & Pt & 0.000 & 199 & Hg & 0.210 \\ 
   200 & Hg & 0.511 & 201 & Hg & 0.391 & 202 & Hg & 0.661 & 203 & Tl & 0.775 \\ 
   204 & Pb & 1.050 & 204 & Hg & 0.102 & 205 & Tl & 0.711 & 206 & Pb & 0.660 \\
   207 & Pb & 0.748 & 208 & Pb & 0.491 & 209 & Bi & 0.066 &     &    &       \\ 
\hline
\enddata
\end{deluxetable*}

The detailed $s$-process fractions 
listed in Table 3 can then be used for disentangling the $s$ and $r$ contributions 
to each isotope, as done, e.g., by \citet{ar9, nikos2}. Since the table contains 
predictions for fractional contributions to each isotope from the $s$-process Main 
Component, the maximum ratio one should get is obviously one, reserved to $s$-only isotopes that do not receive contributions from other components (i.e. those in the atomic mass range between about 90 and about 210). As Table 3 shows, this is not formally true, with the fractional productions slightly differing from unity. One can however
notice that the global value of the variance in the distribution ($\sigma \lesssim$ 15\%) is in the range of general uncertainties known to exist on the product of the two sets of crucial parameters, solar abundances ($N_s$) and cross sections ($\sigma_{N_s}$); the limited residual problems can therefore be simply due to the effects of these 
uncertainties in the input data. It is in any case worth noticing the cases of two special nuclei in Table 3. The first isotope we want to mention is $^{152}$Gd, sometimes indicated as an $s$-only isotope, whose production in our scenario amounts only to about 60\% of its solar concentration.
In fact, we restrain from considering $^{152}$Gd as a real $s$-only isotope, as it is not 
shielded against $\beta^+$ decays and might receive a high contribution from the $p$-process. 
The second nucleus is $^{93}$Nb, which is produced at 66\% of its solar abundance. It is not shielded against $\beta-$decays, so that its production
can be partially due also to the $p$-process, through $^{93}$Mo ($\beta ^+, t_{1/2} \simeq $ 4000
yr) and to the $r$-process, through $^{93}$Zr ($\beta^-$, $t_{1/2} \simeq$ 1.5 Myr). In 
evolved stars its presence is normally seen in objects enriched by mass transfer
from an AGB companion (i.e. in Ba-stars and their relatives, see section 4.2). It must however
be noticed that a further source for $^{93}$Nb production can arise if a Ba-star evolves in 
its turn to the TP-AGB phase, thus undergoing $s$-enrichment for a second time. Such a phenomenon is in fact observed \citep{shetye2}; detailed computations in this last scenario will be necessary
for determining the real $s$-process contribution to $^{93}$Nb.

\section{Constraints from presolar grains and evolved stars}

Surface stellar abundances derived in the previous sections can be then compared 
with various constraints deriving either from spectroscopic observations of evolved 
stars or from the analysis of presolar grains of AGB origin
\citep[primarily of the SiC grain family, see][]{lf95, davis}. For both these
fields a detailed, critical description of the database used would be necessary,
implying rather long analyses that cannot be pursued here. Therefore we
simply anticipate here some relevant examples, postponing more detailed analyses 
to separate contributions.

\subsection{Crucial isotopic ratios in presolar {\rm SiC} grains}

\begin{figure*}[t!!]
\includegraphics[width=0.85\textwidth]{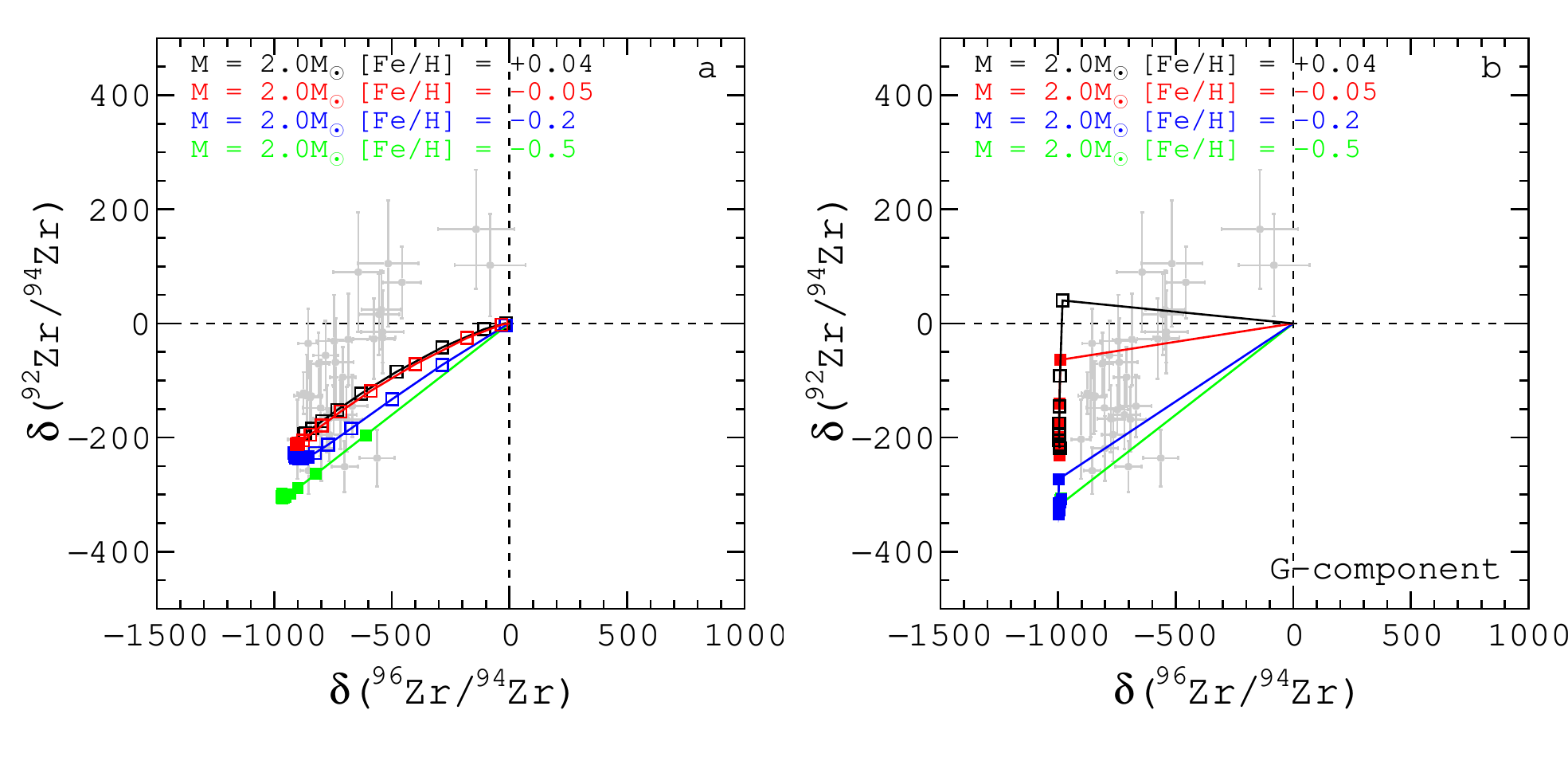}
\caption{The isotopic  ratios $^{92}$Zr/$^{94}$Zr measured in presolar SiC grains, versus $^{96}$Zr/$^{94}$Zr, 
in $\delta$ units, taken from the measurements cited in the text. They are compared with our model sequences for stellar atmospheres (left panel) and for the $G-component$ (right panel), for stars of M = 2 \ms. Lower masses
would not differ much from the plot shown, if not for the more limited extension of the C-rich phase.  
\label{fig:zrdelta2b}}
\end{figure*}

\begin{figure*}[t!!]
\includegraphics[width=0.85\textwidth]{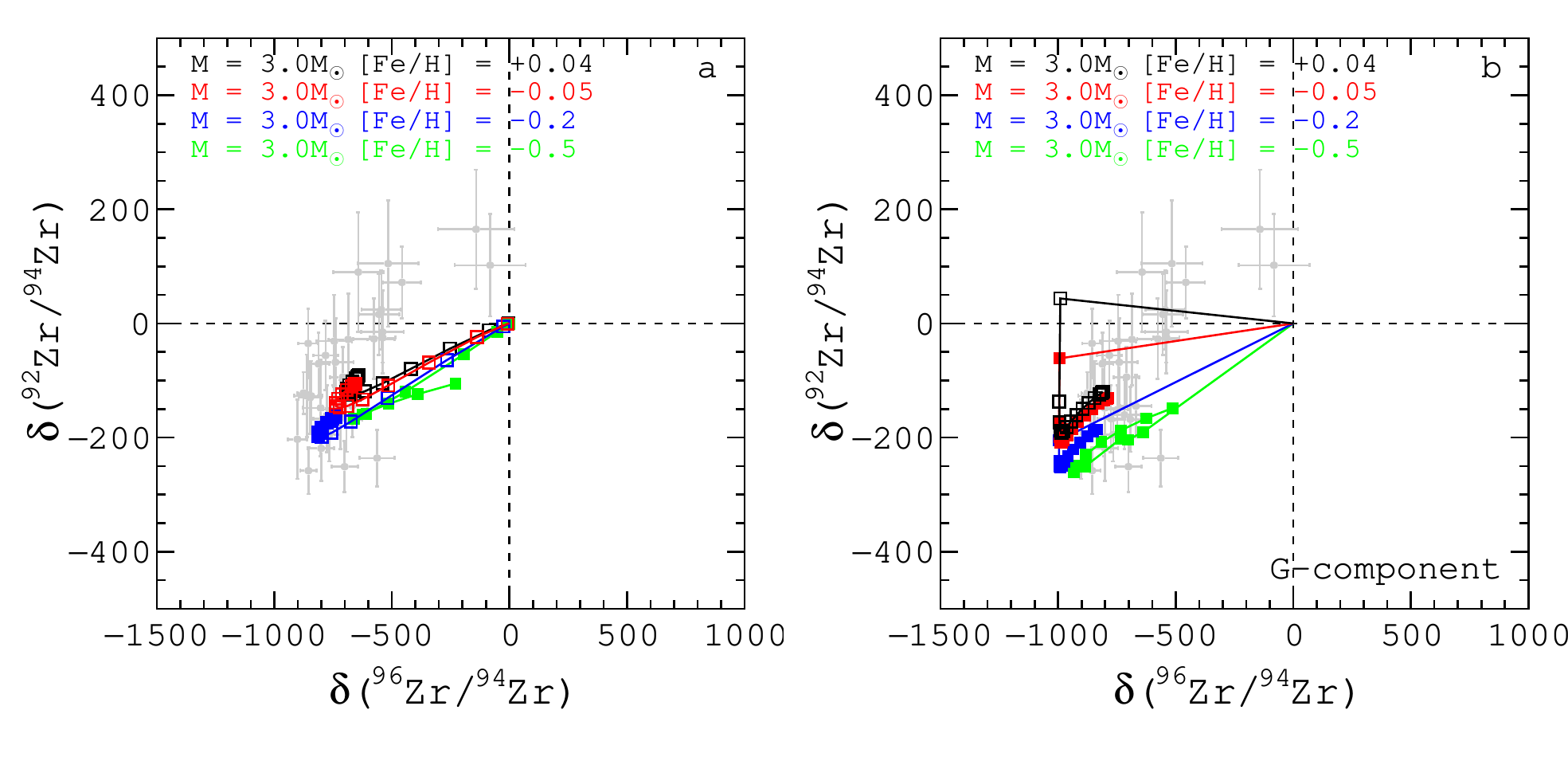}
\caption{Same as in Figure \ref{fig:zrdelta2b}, but for model stars of M = 3 \ms.  
\label{fig:zrdelta2c}}
\end{figure*}

\begin{figure*}[t!!]
\includegraphics[width=0.85\textwidth]{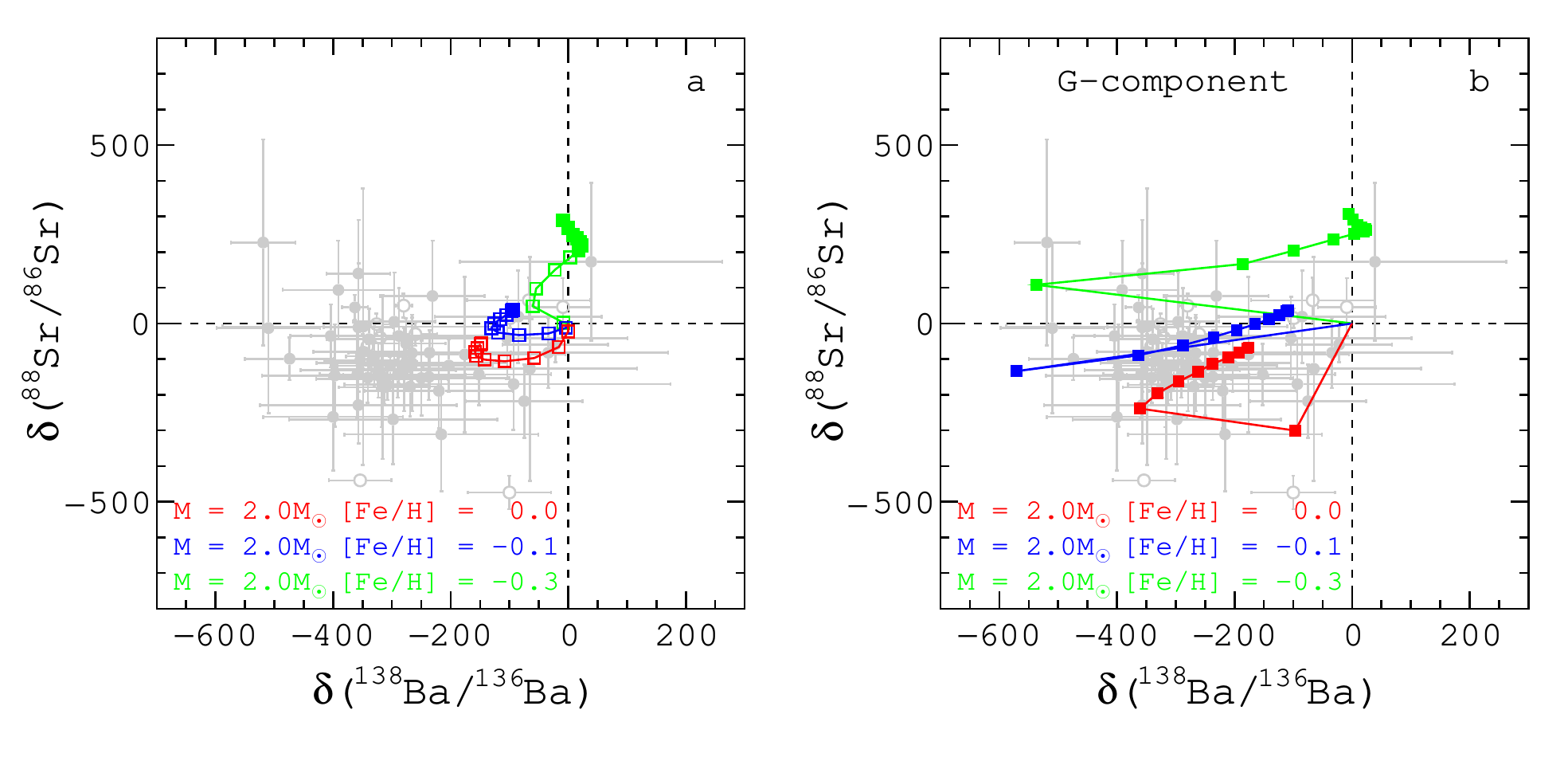}
\caption{The isotopic ratios $^{88}$Sr/$^{86}$Sr observed in presolar SiC grains, versus $^{138}$Ba$^{136}$Ba,
expressed with the $\delta$ notation, taken from the measurements cited in the text, as compared with our model sequences for stellar atmospheres (left panel) and for the $G-component$ (right panel) of stars with M = 2.0 \ms. Again lower mass models do not differ drastically from what is shown. The data are integrated, with respect to \citet{GCA} with those shown by \citet{liu2, stephan}.\label{fig:basrdelta1}}
\end{figure*}

\begin{figure*}[t!!]
\includegraphics[width=0.85\textwidth]{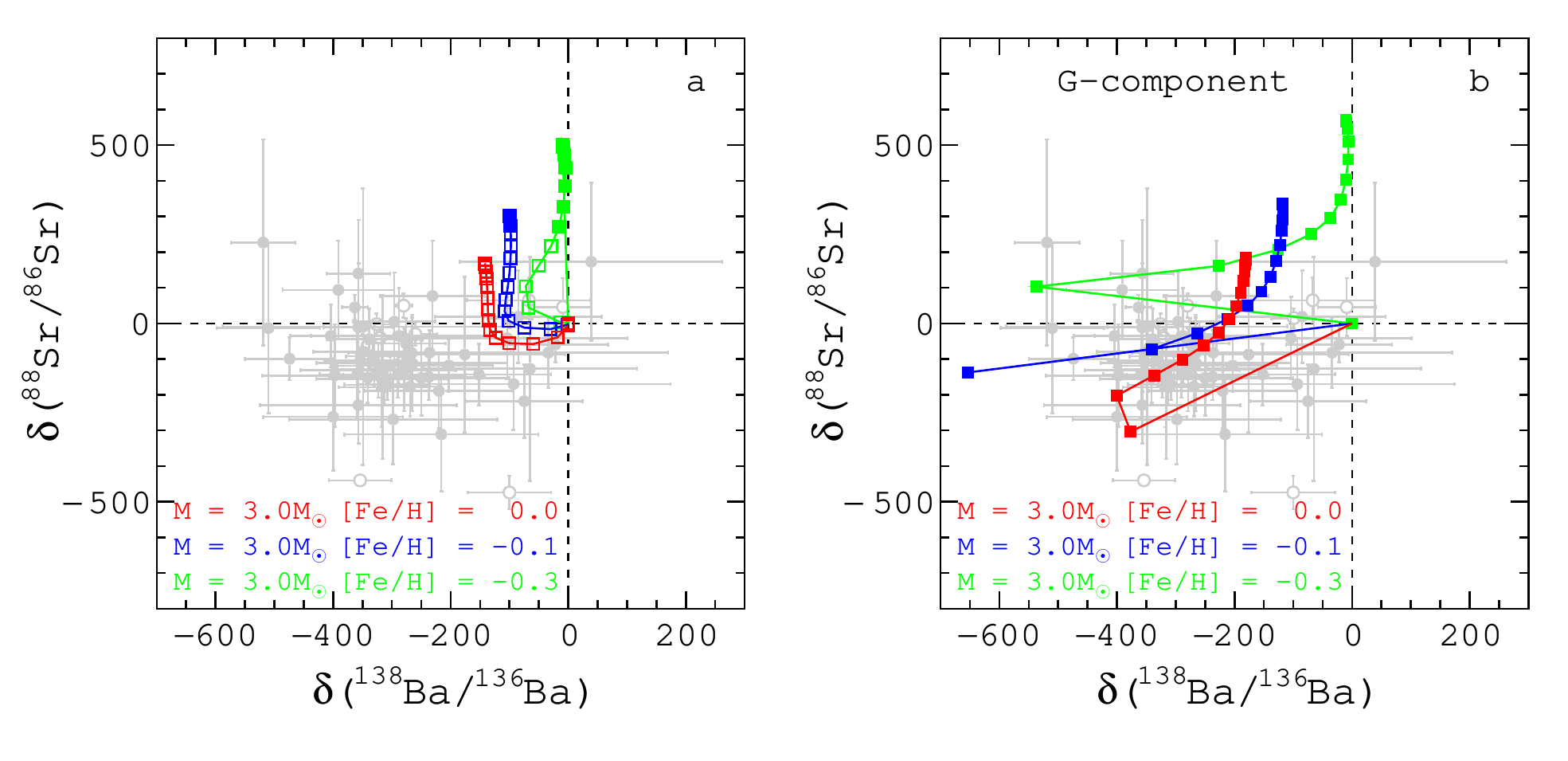}
\caption{Same as in Figure \ref{fig:basrdelta1}, but for model stars of M = 3 \ms.  
\label{fig:basrdelta2}}
\end{figure*}

For presolar grains, results based on a preliminary extension of paper II, 
similar albeit more limited than the present one, were shown in \citet{GCA}.
We refer to that paper for most of the details. Here we want to emphasize that
a few isotopic ratios of trace elements measured recently in presolar 
SiC grains were crucial for excluding some previous parameterized scenarios 
for the formation of the \ctb source. They were also used to suggest that
models based on the indications contained in paper II had instead the 
characteristics required to account for the grain data \citep{liu, liu1}. Such
measurements are therefore important tests to validate nucleosynthesis results 
{deriving} from any specific mixing scheme, as other general constraints on 
$s$-processing are much less sensitive to the details of mixing \citep{bun17}.
It turned out that of particular importance in this respect is the
isotopic ratio $^{92}$Zr/$^{94}$Zr versus $^{96}$Zr/$^{94}$Zr \citep{liu0} 
as well as the correlation between $^{88}$Sr/$^{86}$Sr and $^{138}$Ba/$^{136}$Ba \citep{liu, liu1}. 
These data (expressed in the usual $\delta$ notation) are reproduced in
Figures from \ref{fig:zrdelta2b} to \ref{fig:basrdelta2}, overimposed to our model sequences 
(see discussion below).  

The choice of what to define a carbon-enhanced composition requires some 
comments. SiC grains are carbon-rich solids, so their formation is normal for 
an abundance ratio C/O larger than unity. However, 
recent work on non-equilibrium chemistry in circumstellar
envelopes suggests that the contemporary formation of both O-rich and 
C-rich molecules can occur for wide composition ranges of the environment, 
thanks to various complex phenomena, including the photo-dissociation 
and re-assembly of previously-formed compounds. 
According to \citet{cher}, a
two-component (silicate-carbon) dust may form when the
C/O ratio is sufficiently high, even before it formally
reaches unity, along the AGB evolution. These suggestions
do correspond to the 
observations by \citet{olofsson}, who found carbon-rich compounds (like 
the HCN molecule) in O-rich environments. We therefore allowed a more 
relaxed constraint to the carbon enrichment of the envelope, imposing
that C/O be larger than 0.8.

The composition of the envelope is then largely influenced by mass loss rates,
whose understanding is far from satisfactory. As mentioned before in this note, 
we can avoid being dominated by uncertainties in stellar wind efficiency if we
use, aside to the envelope composition, also the composition of the $G-component$.
{The importance of this last is enhanced by the already-mentioned fact that 
this extrapolated phase mimics well the abundances in flux tubes at each TDU episode, 
hence also that of possible flare-like phenomena disrupting magnetic structures in the
winds and adding C-rich materials to them.}

Figures from \ref{fig:zrdelta2b} to \ref{fig:basrdelta2} are therefore plotted in two panels,
one representing the formal envelope composition when a C-rich situation (C/O $>$ 0.8) has 
been reached, the second showing the pure, C-rich $G$-component. From the figures it is 
clear that, indeed, at least this last offers excellent possibilities of reproducing the 
presolar grain data for critical Sr, Zr and Ba isotopic ratios, which fact confirms the 
viability of the adopted choices for the partial mixing processes generating the \ct-pocket, 
in agreement with the mentioned findings by \citet{liu, liu1}. We notice here that 
accounting for measurements with $\delta$($^{92}$Zr/$^{94}$Zr) $\ge -50$ was considered 
as impossible for $s$-process modeling \citep{lugaro4}. In \citet{liu0, liu} it was underlined how the problem could be alleviated or bypassed through a specific parameterization of 
the \ctb pocket, subsequently recognized to mimic the distribution found in our Paper II \citep{liu1}. Here Figures \ref{fig:zrdelta2b} and \ref{fig:zrdelta2c} specify better 
the reason of this way out. The puzzling isotopic ratios correspond roughly to our $G-component$ 
at the first TDU episodes for high metallicity stars. The explanation of this is straightforward: 
in early cycles of high metallicity models, still at low temperature and having a low neutron density 
in the pulses,  $^{96}$Zr is essentially destroyed, while a marginal production
of $^{92}$Zr occurs. Clearly, the envelope is still O-rich, but for us (as already mentioned) the 
$G-component$ mimics rather well the composition of those flux tubes that, surviving disruption in 
the turbulent convective envelope, reach the surface (as done by the solar magnetic structures 
forming the corona). When a few magnetized blobs reach the atmosphere without mixing, kept 
together by magnetic tension, solids condensed there have a finite probability of preserving a C-enriched composition, being contemporarily slightly $^{92}$Zr-rich and largely $^{96}$Zr-poor. We believe therefore that presolar SiC grains, through those isotopic ratios that are {otherwise} hardly associated to C-rich envelopes,
add a remarkable piece of evidence in favor of our MHD mixing scheme, {naturally accounting  
for the existence C-rich blobs}, where carbon-based dust can be formed even during generally O-rich phases.

In Figures \ref{fig:zrdelta2b} and  \ref{fig:zrdelta2c} the few points at high, positive values of 
$\delta$($^{92}$Zr/$^{94}$Zr) and with $\delta$($^{96}$Zr/$^{94}$Zr) $\ge -500$, which are not covered by 
the region of the models, might be easily fitted, should one consider initial abundance ratios for Zr 
isotopes in the envelope different from solar. This would indeed be the case for stellar models of low metallicity, as the contributions from AGB stars to the lighter isotopes of Zr is lower (see Table \ref{tab:tab3}) than for the reference $^{94}$Zr, implying that the envelope (initial) $\delta$ value for (say) $^{92}$Zr/$^{94}$Zr 
should be slightly higher than zero. We believe, however, that one should restrain from over-interpreting the data: there are, in fact, 
remaining problems in them that hamper too strong conclusions and suggest the need for new 
high-precision measurements. This was extensively discussed in \citet{liu1}, to which we refer 
for details. We also {remind the reader} that important
effects on the Zr isotopes of SiC grains, related to the composition of the parent stars, also invoking 
contributions from super-solar-metallicity AGBs, were suggested by \citet{lugaro8}. This possibility 
and other relevant issues on this subject require to be discussed in more detail in a separate, 
forthcoming paper.

\subsection{Reproducing the abundances of AGB stars and their relatives}

AGBs are in principle precious also because, thanks to the TDU episodes, 
they offer a unique opportunity to observe ongoing nucleosynthesis 
products directly in the producing stars. Several important observational 
studies on their O-rich and C-rich members exist and have been crucial
in revealing their properties and composition \citep{smith1, smith2, smith3, smith4, busso92, b+01, 
abia4, abia82, abia5, abia6, a20,rau, shetye}. However, due to their low temperature and complex 
atmospheric dynamics, when they cross their final (TP) phases they become very 
difficult to observe. At long wavelengths, in their cool circumstellar 
envelopes, $n$-capture elements are easily condensed at relatively low 
temperature \citep{ritch} so that abundances measured for the gaseous phase 
are difficult to correlate with the atmospheric ones. On the other hand, 
these atmospheres are subject to strong radial pulsations, with periods of 
the order of one year (from a fraction to a few), which make them variable 
by several magnitudes (LPVs, or Long Period Variables), see e.g. \citet{wood2}.
They are classified as being Mira or Semi-regular pulsators, with pulsations 
variously dominated by the fundamental or first overtone mode \citep{wood1}. 
Model atmospheres in those conditions are extremely complex 
\citep [see][and citations there]{gautschy, hofner}. Moreover, strong molecular transitions
hamper the observations of crucial elements \citep{abia82}. Owing to these difficulties, 
various families of relatives (having warmer photospheres) have 
become important surrogates in providing information on AGB nucleosynthesis. This includes
in particular binary systems where a surviving companion inherited measurable abundances of
$n$-capture elements through mass transfer by a specific mechanism, called {\it wind accretion}
\citep{jm, bj}. The elements were formed previously in an AGB star now generally evolved to the 
white dwarf phase \citep[see, e.g.][and references quoted therein]{jor1, jor2, escorza1, escorza2, escorza3}.
These objects are often called {\it extrinsic} AGBs \citep{sl, jm}, the most famous class of them being that
of classical Ba-II stars. They were the object of many studies over the years, and of recent 
extensive discussions, with new observational data, see e.g. \citet{jornew}. 
Although we cannot afford such an extended topic in detail here, we need at least to verify our models 
in general on Ba-star constraints. 

Another class of AGB relatives that attracted large 
attention are post-AGB objects that, in their evolution toward the white dwarf stage, cross blueward 
the HR diagram, heating their remaining envelope and passing therefore through various
spectral types corresponding to temperatures warmer than for real AGB stars \citep{vw1, vw2, vw3, vw4,desmedt5,desmedt3}. 
They include extrinsic objects, as well \citep{kama1, kama2}. We need to present at least a couple of
examples of how our model scenario compares with post-AGB constraints. 

\begin{figure}[t!!]
\includegraphics[width=0.85\columnwidth]{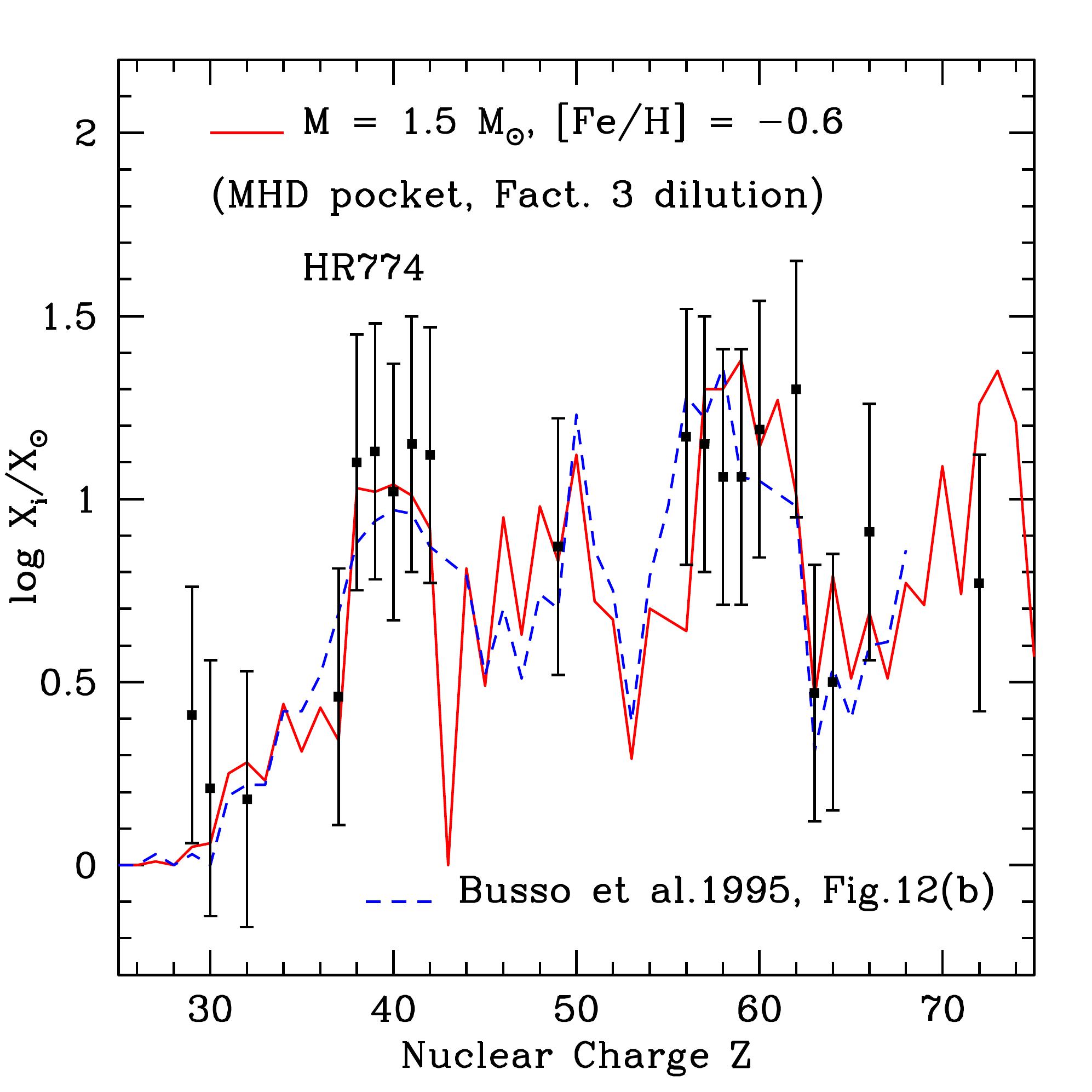}
\caption{Comparison of models with observations for the Ba star HR774. We show
results from the present work (red continuous line) as well as from 
the parametric study by \citet{b+95}, namely their case B (blue dashed line). 
\label{fig:hr774}}.
\end{figure}

\begin{figure}[t!!]
\includegraphics[width=0.85\columnwidth]{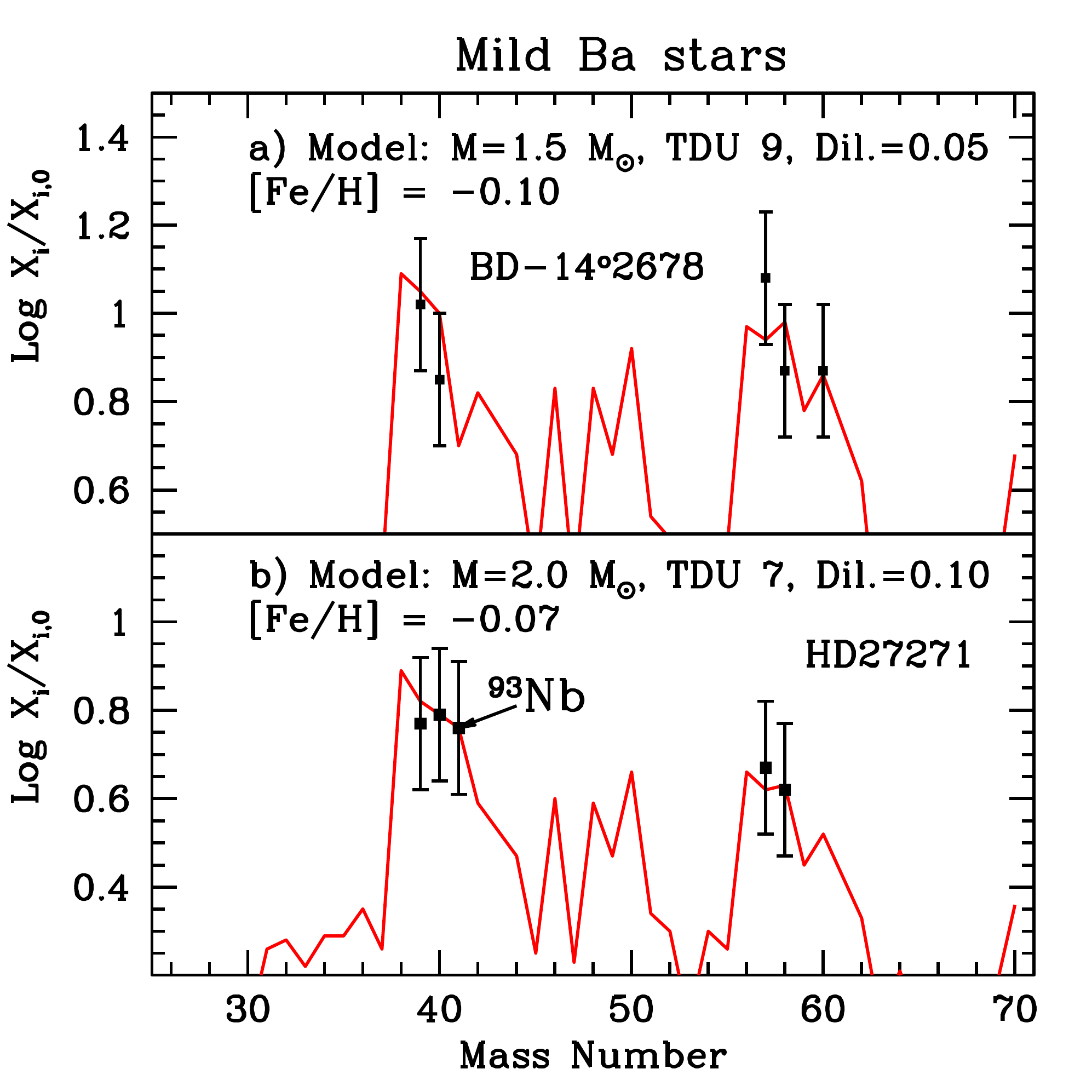}
\caption{Comparison of our models with observations for two mild Ba stars (see text for details) \label{fig:newba}}
\end{figure}

Starting with Ba-stars, extended fits to some of their abundance distributions were done by
us many years ago \citep{b+95, b+01}, as a tool for understanding the extensions of the \ctb
pockets by calibrating parameterized models on observations. An
example of how things have changed now is shown in Figure \ref{fig:hr774}, for HR774 (HD 16458), whose abundances are taken from the measurements by \citet{tl83} and by \citet{s84}. Our original fitting attempt was presented in Table 2 and Figure 12 of \citet{b+95} for two tentative parameterized models. 
Now Figure \ref{fig:hr774} shows the comparison of one such model (model B) with what can be 
obtained in the envelope of a low mass star, of metallicity $[\rm Fe/H] = -0.6$,
having undergone efficient mass transfer from an AGB companion  of about 1.5 \ms. The two model curves, albeit different, are fully compatible with the observed data; what 25 years ago could only be obtained 
by fixing ad hoc the parameters (in particular the abundance of \ctb burnt), is now a natural 
outcome of our scenario with MHD mixing, applied to the mentioned star and without adjusting any further parameter. We might choose several other examples. In most cases, the observations do not include as many
$s$-elements as for HR774, but they have reached a considerable statistical extension and 
are made with more modern instrumentation. Two such examples are shown in Figure \ref{fig:newba}, taken
from the samples by \citet{decastro} and by \citet{jornew}. We choose the star BD$-$14$^o$2678
from the first mentioned list, and HD27271 from the second one. 
In this last case the observations are from \citet{karin}, who also measured the critical element 
Nb, having only one stable isotope, $^{93}$Nb. As mentioned, this nucleus is
produced by decay of the rather long-lived parent $^{93}$Zr and the presence of the daughter 
$^ {93}$Nb is a clear indication that the star is extrinsic. Both the chosen sources are classified as mild Ba-stars by \citet{jornew}.

The examples shown in the figure represent rather typical cases and the quality of the fits is 
in general good. This is however not possible for {\it all} Ba-stars, as some of the data sets 
available contain individual elements with abundances incompatible with any $s$-process 
distribution: e.g. elements belonging to one of the major abundance peaks, which are discrepant 
by large amounts (even one order of magnitude) with respect to their neighbors. The origin of 
these discrepancies is not known: we believe they may be due to difficulties in spectroscopic 
observations. We must in this respect {remind the reader} that the cases here shown {derive} from a very
simple approach to mass accretion (actually, a simple dilution of the AGB material). More
sophisticated treatments exist and should in fact be pursued, as done in some cases in the past
\citep{stan3}.

\begin{figure}[t!!]
\includegraphics[width=0.85\columnwidth]{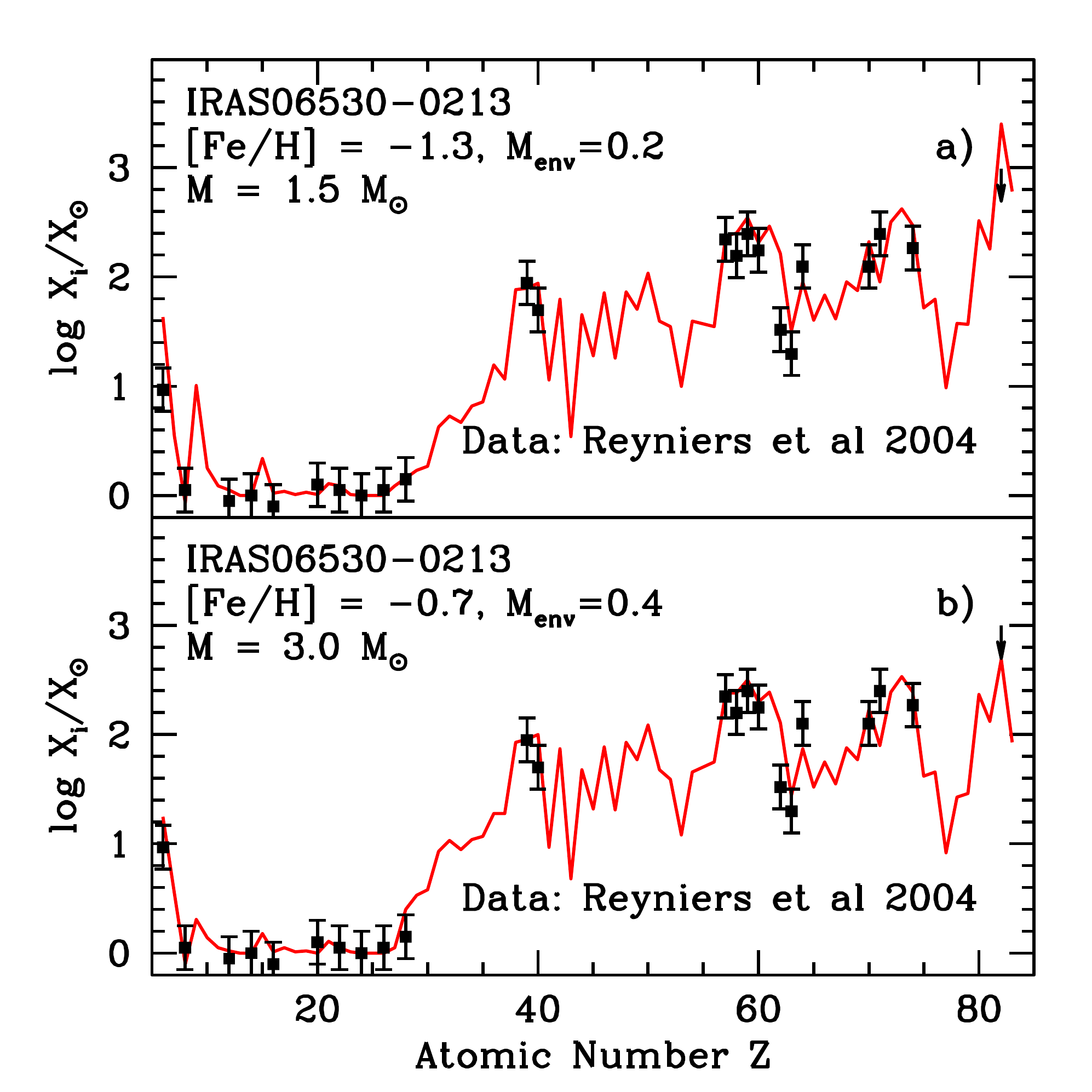}
\caption{Comparison of our model results with the observations for the post-AGB star 
IRAS06530$-$0213. The curve of panel a), from a low mass star, shows the usual discrepancy
on the abundance of Pb, which can in this case be avoided with a fit taken from a more massive star
and a higher metallcity (panel b).\label{fig:IRAS}}
\end{figure}

\begin{figure}[t!!]
\includegraphics[width=0.85\columnwidth]{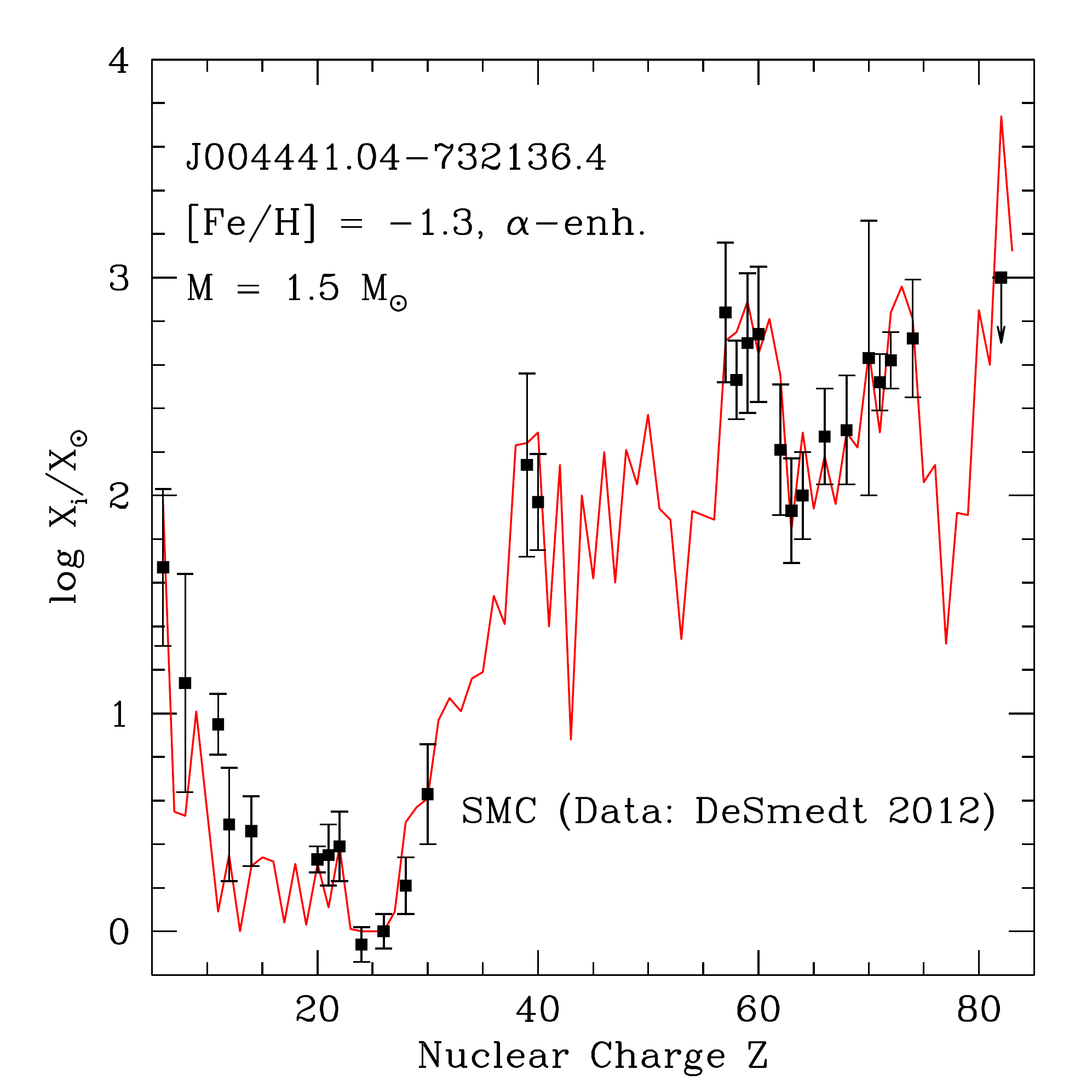}
\caption{Comparison of our model results with the observations for the post-AGB star
J004441.04$-$732136.4. See text for details. 
\label{fig:j00444}}
\end{figure}

An important signature of how $s$-elements are enriched over the galactic history
is provided by the abundance of Pb: its trend is shown here in Figure \ref{fig:pbhsagb}, and roughly
the same behavior was previously found by all groups working in the field: see e.g. \citet{gal1, gm00, lu12}.
As mentioned, the increase in the neutron exposure for lower abundance of the seeds makes it 
unavoidable that, on average, the photospheric abundance expected for Pb increases towards lower 
metallicities. While most observations of metal-poor extrinsic AGB stars
confirmed this evidence, {some of them did not} \citep{ve01, ao01, ve03, be10, bi12}.
{This throwed a shadow on our understanding of the $s$-process scenario, which is, for the rest, robust}.
More doubts on Pb were accumulated in recent years from the second sample of AGB relatives we mentioned, 
i.e. that of post-AGB stars, which again for the rest {confirmed} the known trend of $s$-enrichment
with metallicity \citep{desmedt4}. In the last two decades, various observational studies and detailed
analyses have been performed in this framework. We refer to known works in this field, like e.g.
\citet{vw3, vw4, rey7, desmedt5, desmedt3, desmedt2, desmedt1} and to review papers like \citet{vw1, vw2}
for general reference on the subject. Both for Galactic \citep{desmedt2} and for extra-galactic low-metallicity
post-AGB stars \citep{desmedt4} the expected strong enhancement of Pb was found not to be compatible with
the upper limits determined observationally. {Should these indications be confirmed, in our models the Pb 
problem would} remain, as shown for example in figure \ref{fig:IRAS}, panel a), for observations by \citet{vw4}. It can be sometimes avoided if we refer to stellar models of 3 \msb (see panel b) as, in our scenario, 
the $s$-process efficiency decreases in general for increasing stellar mass. However, the star shown seems 
to be of lower mass and metallicity than found in this purely formal solution \citep{hr03}. We notice, in any case,
how Figure \ref{fig:IRAS} shows that the remaining abundance distribution is reproduced quite well, at 
the same level before possible only through models with parameterized extra-mixing. A similar situation 
emerges from the comparison in Figure \ref{fig:j00444}, done with the observations of the low-mass post-AGB star
J004441.04$-$732136.4 in the Small Magellanic Cloud (SMC) by
\citet{desmedt5, desmedt4}. This star is also classified as a Carbon Enhanced Metal Poor star, enriched in both 
$r$- and $s$-elements  \citep[${\rm CEMP}r$-$s$, see][]{yu-ga}. We recall how these objects often pose very 
difficult problems to detailed modeling of their abundances \citep{stan4}.  {
Very recently, new HST ultraviolet observations of three metal-poor stars by \citet{roed20}, using for 
the first time the Pb II line at $\lambda = $ 220.35 nm, yielded much higher abundances for Pb (by 0.3 -- 
0.5 dex) than previously found with Pb I lines. The authors suggest that these last may lead to 
underestimate the Pb abundance. Although it is premature to derive final conclusions from such suggestions, 
they may open the road for solving a long-lasting discrepancy between observations 
and nucleosynthesis computations for AGB stars and their relatives, thus reconciling also our models with the measured data}. 

\section{Conclusions}

The results of this work can be summarized briefly by saying that we showed how a general scenario for
the activation of the \ctb neutron source in AGB stars can be built on the simple hypothesis
that the required mixing processes {derive} from the activation of a stellar dynamo, in which
an exact, {\it particular} solution to MHD equations is possible on the basis of the simple, 
but plausible average field geometry suggested by \citet{nb}.

In particular, we have shown how, based on that hypothesis, one can avoid all further free 
parameterizations and deduce rather general rules for deriving the extensions and shapes of 
the \ctb distributions left in the He-intershell zone at each TDU episode. Such distributions
provide nucleosynthesis models suitable to explain the known observational constraints on 
$s$-processing. These last include the average $s$-element distribution in our Solar System,
as well as the peculiarities emerging from the isotopic ratios of trace elements measured
in presolar SiC grains. We show in this respect that some such ratios, previously hardly
accounted for by $s$-process models, can be naturally explained if the cool winds of evolved 
low mass stars contain unmixed blobs of materials, transported by flux tubes above the 
convective envelope, as occurring in the Sun. {This hypothesis provides an approximate 
interpretation for the so-called {\it G-component} of AGB $s$-processing.}

Our results also imply a scheme for the enrichment of neutron-capture elements 
in the Galaxy that accounts for most abundance observations of evolved low- and intermediate-
mass stars and has the characteristics previously indicated in the literature as required
for understanding the enhanced heavy-element abundances of young open clusters.

As a consequence of our rather long reanalysis of the mixing processes required for
making the neutron source \ctb available in the late evolutionary stages of red giant stars,
initiated with the studies by \citet{b+07}, and considering that the observational
confirmations so far accumulated give us a sufficient guarantee of robustness, the mixing
scheme developed in the past years has now been released for direct inclusion in full
stellar models of the FUNS series \citep{cris0, cris2}. A first attempt for implementing this integration was recently published
separately by \citet{diego}. 

\bibliography{biblio1}{}
\bibliographystyle{aasjournal}
\end{document}